\documentclass{aa}

\usepackage{graphicx}
\usepackage[colorlinks=true, citecolor=blue, linkcolor=blue, breaklinks=true]{hyperref}
\makeatletter
\renewcommand*\aa@pageof{, page \thepage{} of \pageref*{LastPage}}
\makeatother

\usepackage{xcolor}

\usepackage{txfonts}
\usepackage{bm}

\begin{document}

   \title{The AIDA-TNG project. Abundance, radial distribution, and clustering properties of halos in alternative dark matter models}

   \author{M. Romanello
          \inst{1,2}
          \and
          G. Despali
          \inst{1,2,3}
          \and
          F. Marulli \inst{1,2,3}
          \and
          C. Giocoli\inst{2,3}
          \and
          L. Moscardini\inst{1,2,3}\and M. Vogelsberger\inst{4}     
          }

   \institute{Dipartimento di Fisica e Astronomia “A. Righi” - Alma Mater Studiorum Università di Bologna, via Piero Gobetti 93/2, 40129 Bologna, Italy
   \and
   INAF - Osservatorio di Astrofisica e Scienza dello Spazio di Bologna, via Piero Gobetti 93/3, 40129 Bologna, Italy 
   \and 
   INFN - Sezione di Bologna, Viale Berti Pichat 6/2, 40127 Bologna, Italy
   \and 
   Department of Physics, Kavli Institute for Astrophysics and Space Research, Massachusetts Institute of Technology, Cambridge, MA 02139, USA
\\\\
        \email{massimilia.romanell2@unibo.it}
             }

   \date{}

 
  \abstract  
  {
  Warm and self-interactive dark matter cosmologies have been proposed as nonbaryonic solutions to the tensions between the $\Lambda$ cold dark matter model and observations at the kiloparsec scale. In this paper, we used the dark matter-only runs of the \textsc{aida-tng} project, a set of cosmological simulations of different sizes and resolutions, to analyze the macroscopic impact of alternative dark matter models on the abundance, radial distribution, and clustering properties of halos. We adopted the halo occupation distribution formalism to characterize the evolution of its parameters $M_1$ and $\alpha$ with the mass and redshift selection of our sample. By dividing the halo population into centrals and satellites, we were able to study their spatial density profile. We found that a Navarro-Frenk-White model is not accurate enough to describe the radial distribution of subhalos and that a generalized Navarro-Frenk-White model is required instead. Warm dark matter models, in particular, present a cuspier distribution of satellites, whereas self-interacting dark matter exhibits a shallower density profile. Moreover, we found that the small-scale clustering of dark matter halos provides a powerful tool for distinguishing among alternative dark matter scenarios, in preparation for a more detailed study that fully incorporates baryonic effects and for a comparison with observational data from galaxy clustering.
}


   \maketitle

\section{Introduction}
In the past decades, a significant part of the cosmological investigation has been dedicated to shedding light on the central topic of the origin and the growth of cosmic structures. These efforts led to the formulation of the so-called hierarchical model of structure formation \citep{White78}. According to this, structure formation is the result of the growth of primordial overdensities in the matter distribution that arise at the end of the inflationary era from initial quantum fluctuations in the early Universe. In this framework, gravitationally bound systems, such as galaxies and galaxy clusters, are the product of gravitational infall, the progressive merging of dark matter halos, and the catching up of baryons \citep{White78, Bardeen86, Tormen98}. \\
\indent Several physical processes contribute to the formation and assembly history of these complex systems. Among them, the long-range nonlinear gravitational interaction sets the dynamic of the overall matter distribution, while different radiative processes, including stellar emission and gas cooling, affect the baryonic matter in different phases. Finally, a series of feedback phenomena, such as supernova explosions \citep{Efstathiou2000, Smith18}, stellar winds \citep{Springel03, Dave11, Hopkins12}, and active galactic nucleus emission \citep{Bower06, Weinberger18}, reshape the matter distribution as well as the star formation rate \citep{Scharre24}. Due to the large number of interactions and physical processes involved, an analytical treatment of structure formation is currently unfeasible.\\
\indent For this reason, numerical simulations have been developed in a wide range of sizes and resolutions that target well-defined scientific cases. They offer important tools for investigating the statistical properties of cosmological systems and testing models that will be applied to observations and real datasets. In particular, for a given set of cosmological parameters, N-body simulations trace the evolution of the dark component alone, following the gravitational interaction of dark matter particles over cosmic time 
\citep[][\citeyear{Springel05}]{Springel01}, while hydrodynamical simulations also include baryons, allowing us to understand the complex interplay between their different components, such as gas, stars, and black holes \citep{Duffy10, Vogelsberger14Nature, Chisari18, Springel18, Vogelsberger20}. \\
\indent The clustering of large-scale structure (LSS) is recognized today as a fundamental cosmological probe because during different periods of the cosmic expansion, it keeps the memory of the footprint left by the initial conditions of the Universe. In practice, this offers us the possibility to provide constraints on the main cosmological parameters, such as the matter density content of the Universe and the normalization of the primordial fluctuations \citep{Hildebrandt20, Asgari21, Abbott22, Amon22, Lesci22, Secco22, Romanello24}. In addition, the clustering of LSS is typically combined with other cosmological probes, such as number counts of cosmic structures \citep{To2021}, or a simultaneous study of different two-point statistics \citep{Heymans21}, which includes cosmic shear \citep{Kilbinger15, Hildebrandt20, Asgari21} and galaxy–galaxy lensing, and can potentially highlight differences in the concordance cosmological paradigm.\\
\indent Currently, the most widely acknowledged cosmological scenario is the $\Lambda$ cold
dark matter ($\Lambda$CDM) model, which assumes that dark matter particles are in a ‘cold' version, namely in the form of very massive nonrelativistic candidates, such as weakly interacting massive particles, with a mass $m_X>1$ GeV, or condensates of light axions, with $m_X\lesssim 10^{-3}$ eV \citep{Strigari13}. The $\Lambda$CDM model is consistent with observations at various scales that range from the typical intergalactic distances to the size of the cosmological horizon. It also provides robust
explanations for observations of the cosmic microwave background, for the abundances of
light elements as products of the primordial nucleosynthesis, and for the accelerating expansion of the Universe. Moreover, it reasonably describes the distribution and statistical properties of the LSS. \\
\indent However, some possible tensions have been suggested by observations on the galactic and sub-galactic scales, of the order
of kiloparsec \citep[see][for reviews]{Weinberg15, Bullock17, DelPopolo17}. Among them, the ‘missing satellites' \citep{Klypin99} and the ‘too-big-to-fail' problems \citep{Boylan-Kolchin11}, which are related to the fact that the abundance of satellites around the Milky Way does not match the number of galactic subhalos measured in N-body simulations, and the ‘cusp-core' problem, which arises because the observed halo density profiles are less concentrated than the predicted cusped Navarro-Frenk-White profiles \citep[NFW,][]{Navarro96}. Although a recent galactic census has significantly alleviated the missing satellite problem \citep{Kim18}, the issue remains open scientifically. Usually, baryonic phenomena are invoked as possible solutions \citep{Brooks13}; for example, photoionization feedback to quench star formation in low-mass halos \citep{Busha10, Castellano16, Yue18}, and supernova feedback and tidal stripping to expel gas from the galactic environment, resulting in a redistribution of matter that erases the internal cusps of density profile and in a reduction of the number of faint satellites \citep{Brooks13, Bullock17}. \\
\indent Another perhaps more intriguing possibility is the exploration of the dark sector by investigating the macroscopic consequences of alternative cosmological scenarios that postulate the existence of warm dark matter \citep[WDM,][]{Bode2001, Viel05, Schneider2012, Lovell14} or self-interacting dark matter particles \citep[SIDM,][]{Spergel2000, Vogelsberger12}. In the first case, the growth of cosmic perturbations smaller than a characteristic free streaming length is suppressed by the thermal motion of collisionless dark matter particles. This naturally produces shallower density profiles and fewer low-mass halos \citep{Menci16}. This observationally translates into a turnover at the faint end of the rest-frame galaxy UV luminosity function \citep{Corasaniti17} and in a general delay of galaxy formation \citep{Dayal15, Menci18}, with a possible effect on the timeline of the reionization process \citep{Dayal17, Carucci19, Romanello21}.
In the latter case, dark matter particles have a nonzero cross section of self-interaction, $\sigma$, which yields a non-negligible scattering probability that allows for a transfer and a redistribution of energy and momentum between different parts of dark matter halos \citep[see][for reviews]{Tulin18, Adhikari22}. As a result, the inner region of the density profile tends to become cored and isothermal, whereas the outer part, in which the density and the interaction rate are lower, is well described by an NFW profile. In more realistic SIDM models, self-interaction is determined by a massive mediator and $\sigma$ decreases with the relative velocities of the interacting particles \citep{Vogelsberger12, Correa21, Adhikari22}. Systems such as galaxy clusters, which have a large velocity dispersion, are therefore less affected than faint galaxies. As a consequence, SIDM affects on kiloparsec rather than on megaparsec scales, which agrees with what is expected from the observational evidence in the local Universe. \\ 
\indent This publication is part of a series of works that make use of a suite of cosmological simulations \citep{Despali25} to study and characterize the properties of dark matter halos, such as their shapes \citep{Giocoli25}, gas distribution \citep{Zhang26}, concentrations and density profiles \citep{Despali25b}, in alternative dark matter models. The management of halo catalogs, the measurements of the relevant two-point statistics, the cosmological modeling, and the Bayesian inference presented in this study were performed with \textsc{CosmoBolognaLib}\footnote{\url{https://gitlab.com/federicomarulli/CosmoBolognaLib}} \citep{MarulliCBL}, a set of free software libraries for C++ and Python.
\section{The AIDA-TNG simulations}
\label{sect_AIDA}
\begin{table*}[h]
\caption{Overview of the dark matter-only runs of the \textsc{aida-tng} simulations, including box size, dark matter particle mass, and available dark matter models.}
\centering
\resizebox{0.9\textwidth}{!}{%
\begin{tabular}{|c|c|c|c|c|c|c|c|c|}
\hline 
Name & Box side [Mpc $h^{-1}$] & $m_{\text{DM}}$ [$\mathrm{M}_{\odot}$ $h^{-1}$] & CDM & WDM 1 keV & WDM 3 keV & WDM 5 keV & SIDM $1 \mathrm{~cm}^2 \mathrm{~g}^{-1}$ & vSIDM \\
\hline 
100/A & 75 & $4.8 \times 10^7$ & $\checkmark$ & - & $\checkmark$ &  - & $\checkmark$ & $\checkmark$ \\ 
\hline 
50/A & 35 & $2.9 \times 10^6$ & $\checkmark$ & - & $\checkmark$ & $\checkmark$ & $\checkmark$ & $\checkmark$ \\ 
\hline 
50/B & 35 & $2.3 \times 10^7$ & $\checkmark$ & $\checkmark$ & $\checkmark$ & - & $\checkmark$ & $\checkmark$ \\ 
\hline 
\end{tabular}%
}
\vspace{5pt} 
\label{tab_boxes}
\end{table*}
\begin{figure}[htbp!]
\centering
\includegraphics[width=0.5\textwidth]{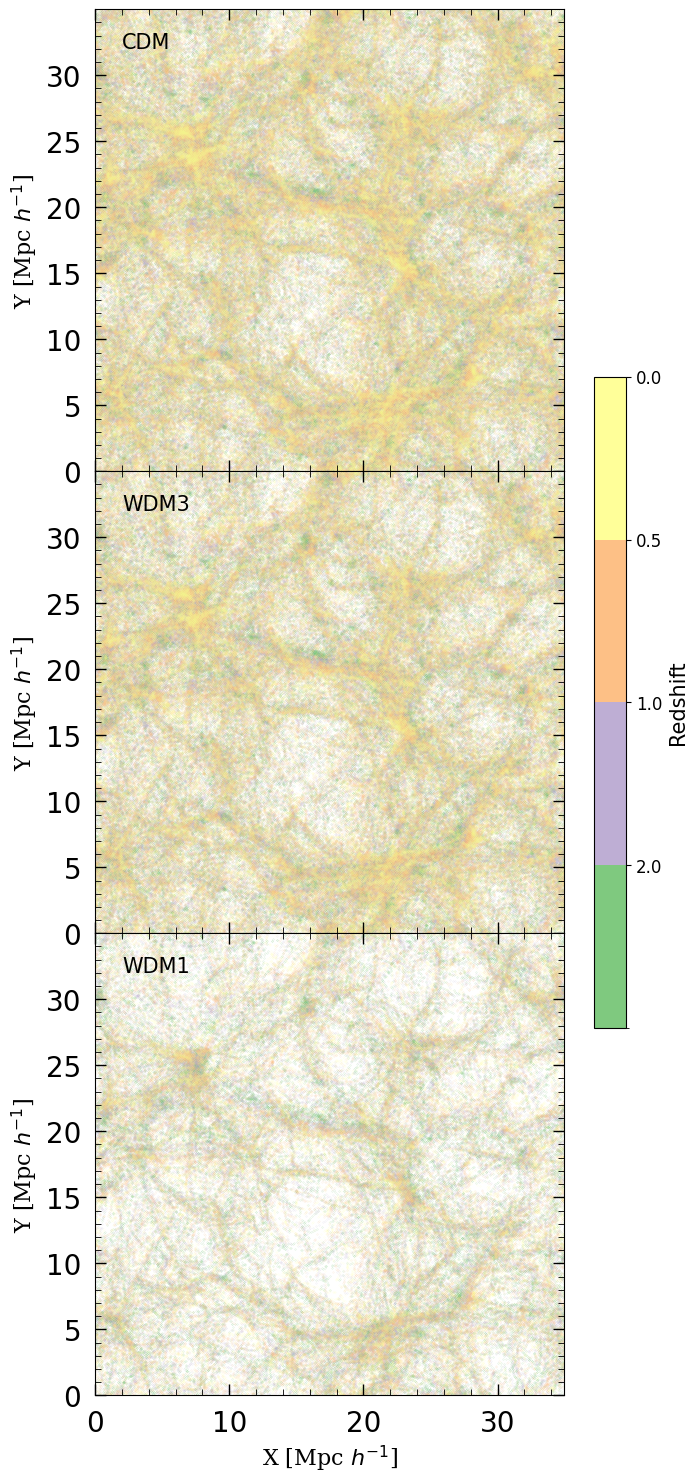}
\caption{From top to bottom: Spatial distribution of dark matter halos in box 50/B for the CDM, WDM3, and WDM1 cosmologies. The color bar shows the corresponding redshift in the range $0<z<2$. }
\label{fig_simulation}
\end{figure}
The \textsc{aida-tng} sample \citep[Alternative Dark Matter in the TNG universe,][]{Despali25} is a set of cosmological simulations designed to study the effect of alternative dark matter models (WDM and SIDM) on multiple scales. Boxes of $110.7$ and $51.7$ Mpc on a side, corresponding to $75$ and $35$ Mpc$\,h^{-1}$, are presented in \citet{Despali25}; we refer to them as "100/" and "50/". Each box is simulated in up to six dark matter scenarios, as well as with and without baryons, and labeled according to its decreasing mass resolution: $m_\mathrm{DM} = 4.8\times 10^7$ M$_\odot \, h^{-1}$ for 100/A, $m_\mathrm{DM} = 2.9\times 10^6$ M$_\odot \, h^{-1}$ for 50/A, and $2.3\times 10^7$ M$_\odot \, h^{-1}$ for 50/B, respectively. The simulations are listed in Table \ref{tab_boxes}, with respective details and available cosmologies. The baryonic physics is included through the TNG galaxy formation model \citep{Weinberger17, Pillepich18}. \\
\indent The starting redshift of the simulations is $z=127$, with initial conditions derived with the Zel'dovich approximation, as also in the IllustrisTNG simulations \citep{Vogelsberger14, Pillepich2018}. The  friends-of-friends (\textsc{FoF}) and \textsc{subfind} algorithms \citep{Springel2001} have been used to identify dark matter halos and subhalos, respectively. First, the \textsc{FoF} group finder
is run. We then refer to subhalos as sets of locally overdense, gravitationally bound particles that are hosted in each \textsc{FoF} detection. The position of central subhalos coincides with the minimum of the gravitational potential, that is, with the centers of the \textsc{FoF} halo \citep{Pillepich2018}. Therefore, throughout this paper, we consider satellites and subhalos to be synonyms.\\
\indent For the purpose of our work, we focus on dark matter-only 50/A and 50/B boxes at $z=2$, $z=1$, $z=0.5$ and $z=0$, in order to take advantage of the resolution and of the largest variety of dark matter models, while leaving further study on the baryonic influence to the future. The cosmological parameters of the simulations are taken from \citet{Planck16}: $\Omega_\mathrm{m} = 0.3089$, $\Omega_\Lambda = 0.6911$, $\Omega_\mathrm{b} = 0.0486$, $H_0 = 0.6774$, and $\sigma_8 = 0.8159$. \\
\indent The \textsc{aida-tng} project explores the same cosmological volumes in different alternative dark matter scenarios. In particular, for WDM cosmologies, dark matter is considered a ‘thermal relic', resulting from the freeze-out of a dark component in thermal equilibrium with the early Universe. From a physical point of view, the free streaming of dark matter particles determines a suppression in the power spectrum, $P(k)$, which can be modeled through the transfer function, $T(k)$ \citep{Bode2001, Viel05, Schneider2012}, 
\begin{equation}
\label{eq_transfer_function}
   T(k) \equiv \sqrt{\frac{P_{\mathrm{WDM}}(k)}{P_{\mathrm{CDM}}(k)}}  =\left[1+(\epsilon k)^{2 \nu}\right]^{-5 / \nu}, 
\end{equation}
where $\nu=1.2$, and 
\begin{equation}
    \epsilon  = 0.048\left(\frac{\Omega_X}{0.4}\right)^{0.15}\left(\frac{h}{0.65}\right)^{1.3}\left(\frac{\mathrm{keV}}{m_X}\right)^{1.15}.
\end{equation}
Here, $\Omega_X$ is the density parameter of the dark matter. 
We consider three thermal WDM cosmologies, with particle mass corresponding to $m_X=1-3-5$ keV, respectively. Lighter models such as WDM1 and WDM3, with $m_X=1-3$ keV, have been ruled out by Lyman-$\alpha$, number counts, UV luminosity function observations \citep{Viel13, Menci16, Corasaniti17, Dekker22, Villasenor23} and are used only for a broader characterization of WDM abundances and clustering. Furthermore, in WDM cosmologies, the halo mass function experiences a suppression given by \citep{Lovell14}
\begin{equation}
\label{eq_suppression_hmf}
    \frac{n_{\mathrm{WDM}}}{n_{\mathrm{CDM}}}=\left(1+\gamma \frac{M_{\mathrm{hm}}}{M}\right)^\beta,
\end{equation}
where $\gamma=2.7$, $\beta=-0.99$ and $M_\mathrm{hm}$ is the half-mode mass, namely, the mass at which the transfer function in Eq. \eqref{eq_transfer_function} is equal to 1/2. \\
\indent We consider SIDM models with a constant cross section (SIDM1, with $\sigma/m_X = 1$ cm$^2$ g$^{-1}$) and, more realistically, with the velocity-dependent cross section presented in \citet[][vSIDM]{Correa21}, which rapidly declines at the typical velocity dispersion of massive bound systems. The model fits to observations of central densities of dwarf spheroidal galaxies, and it is based on a self-interaction described by a particle mass $m_X = 53.933 \pm 9.815$ GeV and a massive mediator in a Yukawa potential, with mass $m_\varphi = 6.605 \pm 0.435$ MeV \citep{Correa21}.\\
\indent In Fig. \ref{fig_simulation} we show the spatial distribution of dark matter halos in box 50/B at different redshifts and cosmologies. The nodes and filaments become progressively clumpier toward low $z$, as they collect matter from the background. In particular, WDM models show a clear progressive suppression in the number density of bound structures. SIDM cosmologies, on the other hand, are not represented, since, upon qualitative inspection, they do not exhibit a significant difference with respect to $\Lambda$CDM.  
\section{The halo occupation distribution model}
\label{sect_HOD}
Discrete structures, such as galaxies and dark matter halos, arise at the overdensity peaks of the Universe; therefore, they are biased tracers of the underlying matter distribution. On large scales, this reduces to a linear, scale-independent relation between the halo and the dark matter density contrast fields, $\delta_{\mathrm{h}}(\boldsymbol{x})$ and $\delta(\boldsymbol{x})$, through a bias factor, $b_{\mathrm{h}}(z)$, 
\begin{equation}
\label{eq_biased_density}
  \delta_{\mathrm{h}}(\boldsymbol{x})= b_{\mathrm{h}}(z)\delta(\boldsymbol{x}). 
\end{equation}
On smaller scales, however, it is challenging to accurately model the clustering of halos because nonlinear effects emerge.\\
\indent In recent years, the halo occupation distribution (HOD) formalism has been widely used in number count and clustering studies because it represents a simple and powerful theoretical tool for approximating galaxy abundances, bias, and small-scale galaxy clustering \citep{Zheng05, Giocoli10, Asgari23}. It is based on the halo model, namely the idea that all the matter in the Universe is collected into halos \citep{Cooray2002}. Under this hypothesis, the question comes down to understanding how dark matter halos cluster together, based on their classification as central or satellites, and how luminous tracers, such as galaxies, correspondingly populate them.
\subsection{Number counts}
\label{sec_hod_number_counts}
\indent According to the HOD model, the average number of halos is given by 
\begin{equation}
    \label{eq_occupation_number}
    \langle N_h \rangle = \langle N_c \rangle + \langle N_s \rangle, 
\end{equation}
where $\langle N_c \rangle$ is the average number of central halos, while $\langle N_s \rangle$ represents the average number of satellites. Unlike galaxy clustering, where $\langle N_c \rangle$ is parameterized according to the details of the galaxy formation process and of the uncertainties in the luminosity-mass relation, for dark matter halos it is simply given by a Heaviside step function \citep{Zehavi05}, 
\begin{equation}
   \langle N_c \rangle = \Theta(M-M_\mathrm{min}),
\end{equation}
where $M_\mathrm{min}$ is the mass cut of the sample, below which the number of central and satellite halos is equal to zero. There must be one central halo in order to have satellite halos, and thus, for masses higher than $M_\mathrm{min}$, where $\langle N_c \rangle=1$, we can write the number of subhalos as follows: 
\begin{equation}
    \label{eq_number_satellites}
    \left\langle N_{s}\right\rangle=\langle N_c \rangle \left(\frac{M-M_0}{M_1}\right)^\alpha, 
\end{equation}
where $M_1$ and $\alpha$ are the normalization and slope of the power law, respectively, while the parameter $M_0$ is included in more general HOD models, as a satellite truncation mass, which is different from $M_\mathrm{min}$. As we also find in our analysis, however, it is generally poorly constrained (\citealp{Watson10}, \citeyear{Watson12}; \citealp{Piscionere15}; \citealp{Skibba15}).
\\
\indent In this sense, the halo occupation number, $\langle N_h \rangle$, gives us a first indication of the HOD parameters. Moreover, it allows us to test whether the distribution of satellites within halos of a given mass follows Poissonian statistics. In particular, we can investigate the average number of pairs associated with the same halo, which brings clustering information and is the base for the one-halo term of the two-point correlation function \citep{Zheng05},    
\begin{equation}
\langle N_h(N_h-1)\rangle= 2\langle N_{c} N_{s}\rangle +\langle N_{s}\left(N_s-1\right)\rangle = \langle N_{s}^2\rangle+\langle N_{s}\rangle. 
\end{equation}
For a Poissonian distribution, this reduces to $\langle N_h(N_h-1)\rangle=\langle N_{s}\rangle^2+2\langle N_{s}\rangle$. Therefore, for massive halos that contain satellites, the ratio $\langle N_h(N_h-1)\rangle/\langle N_h\rangle^2\xrightarrow{}1$, as we discuss in Sect. \ref{sec_number_counts}.
\subsection{Radial distribution of subhalos}
\label{sec_model_density_profile}
\indent It is often assumed that the distribution of tracers follows the distribution of dark matter. For this reason, an NFW profile \citep{Navarro96} is generally adopted in clustering studies. However, it might not be sufficiently accurate to describe the small-scale distribution of subhalos, introducing systematic uncertainties in the parameter values of $\alpha$ and $M_1$ and in cosmology. In particular, we expect the concentration of satellite halos, $c_h = f_c\, c_{200c}$, to differ from that of dark matter, with $f_c$ as a free parameter of our model\footnote{The quantities with the subscript $200c$ are defined within a sphere where the density is 200 times the critical density of the Universe, $\rho_c(z)$.}. Therefore, following the approach of 
\citet{More09}, \citet[][\citeyear{Watson12}]{Watson10} and \citet{Piscionere15}, we consider a generalized NFW (gNFW),
\begin{equation}
\label{eq_generalized_NFW}
    \rho(r)=\frac{\rho_s}{\left(\frac{r}{r_s}\right)^\gamma\left(1+\frac{r}{r_s}\right)^{3-\gamma}},
\end{equation}
where $\gamma$ corrects the slope of the profile and $r_s=R_{200c}/(f_c\,c_{200c})$ is the scale radius, which varies depending on the halo mass and concentration. To estimate the latter, we use the $c_{200c}-M_{200c}$ relation from \citet{Duffy08}. Finally, $\rho_s = M_{200c}/(4\pi r_s^3\int_0^{R_{200c}/f_c} x^{2-\gamma}/\left(1+x\right)^{3-\gamma}\mathrm{d}x)$ is the characteristic density, where $x\equiv r/r_s$. For example, for a luminous red galaxy sample, the parameter $\gamma$ is found to be an increasing function of luminosity, while $f_c$ is a decreasing one \citep{Watson12}. We retrieve the NFW profile when both $\gamma$ and $f_c$ are equal to one. A value of $f_c$ smaller than one boosts $r_s$, eventually to a radius greater than $R_{200c}$, which means that the density profile keeps its inner slope $-\gamma$ even at larger radial separations, behaving like a pure power law \citep[][\citeyear{Watson12}]{Watson10}. Therefore, in order to explore the full radial profile, the upper limit of the integral is adjusted accordingly. Differences in $\gamma$ and $f_c$ among alternative cosmologies also account for variations of the profile shape and of the relative number of subhalos in the inner part of dark matter groups, respectively. As we show in the next section, this also impacts on the clustering properties of central and satellite halos. 
\subsection{Clustering}
\label{sect_model_twop}
\begin{figure*}[htbp!]
\centering
\includegraphics[width=0.9\textwidth]{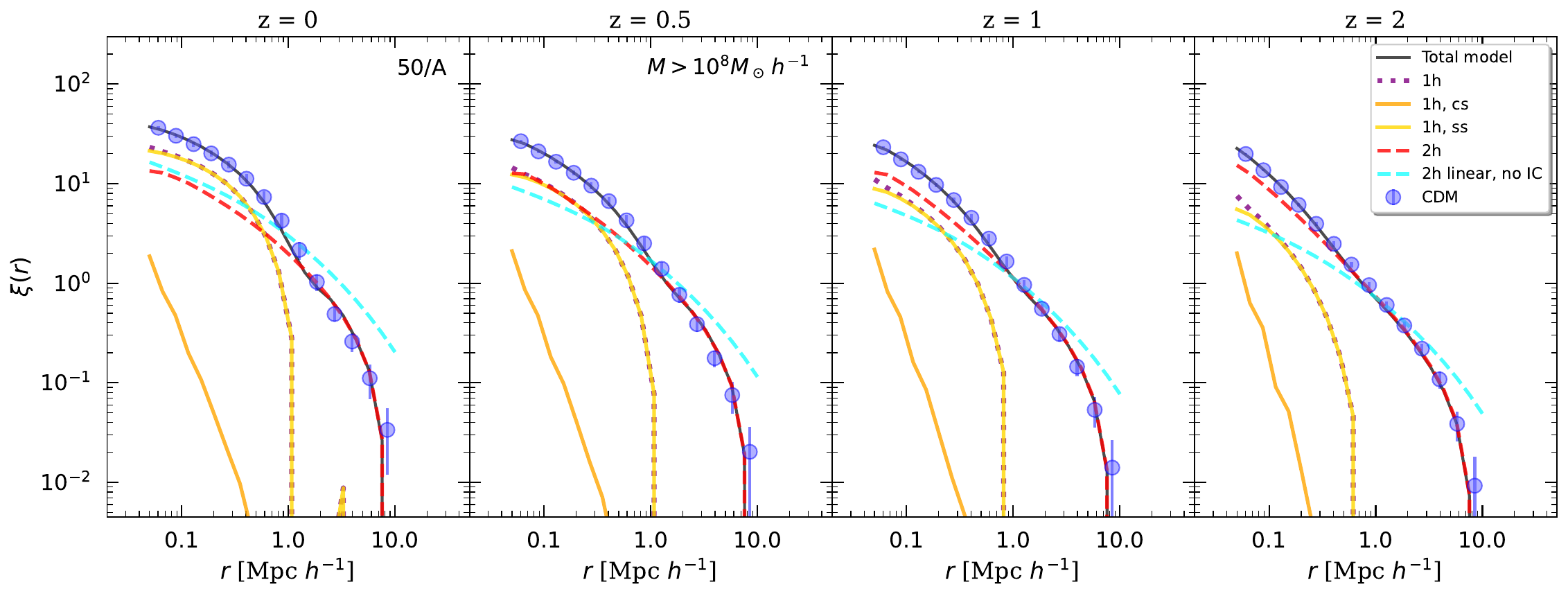}
\caption{From left to right: Different contributions to the two-point correlation function from $z=0$ to $z=2$. The blue circles represent the measurement for box 50/A for CDM cosmology considering halos with $M>10^8$ M$_\odot \, h^{-1}$. The corresponding errors are obtained with the jackknife method. The one-halo term is given by the dotted purple line and it is the sum of $\xi_{cs}$ (solid orange line) and $\xi_{ss}$ (solid yellow line). The linear two-halo term (cyan dashed line) is then corrected for integral constraint, nonlinearities and halo exclusion (dashed red line). Finally, the solid black line represents the total best-fit model.
}
\label{fig_xi_model}
\end{figure*}
We use the HOD formalism to model the small-scale clustering of dark matter halos in cold and alternative dark matter scenarios. In particular, the halo model distinguishes two contributions to the correlation function,
\begin{equation}
\label{eq_xi_tot}
    \xi(r)=\xi_{1h}(r)+\xi_{2h}(r),
\end{equation}
where $\xi_{1h}(r)$ is the one-halo term, accounting for the clustering of matter or tracers hosted by the same halo, while $\xi_{2h}(r)$ is the two-halo term, given by the pairs counted between different halos. The same formalism is equivalently written in Fourier space as 
\begin{equation}
\label{eq_pkhalomodel}
    P(k)=P_{1h}(k)+P_{2h}(k). 
\end{equation}
In Eq. \eqref{eq_pkhalomodel}, the former term dominates at small scales, in a full nonlinear range, that is, below the typical size of dark matter halos, while the latter describes large-scale clustering, where the impact of nonlinearities is smaller and $P(k)= b_\mathrm{h}^2P_\mathrm{lin}(k)$. The linear power spectrum, $P_\mathrm{lin}(k)$, is computed according to the approximation of \citet{Eisenstein98}. This occurs for separations greater than the size of the largest dark matter halo. \\
\indent Dealing with two different populations of tracers, in the one-halo term, we should separate the contribution of central-satellite and satellite-satellite pairs to the power spectrum, $P_{cs}$ and $P_{ss}$, respectively, 
\begin{equation}
P_{1h}(k)=P_{cs}(k)+P_{ss}(k). 
\end{equation}
In particular, 
\begin{equation}
    P_{cs}(k, z)=\frac{2}{n_h^2(z)} \int_{M_{\min}}^{M_{\max }}\langle N_c N_s\rangle n(M, z) \tilde{u}_s\left(k, M, z\right) \mathrm{d} M.
\end{equation}
Here, the central and satellite occupation functions are independent, and therefore, $\langle N_c N_s\rangle = \langle N_c \rangle\langle N_s\rangle$. The normalized distribution of satellites within the halo, in Fourier space, is $\tilde{u}_s\left(k, M, z\right)$, directly derived by numerically transforming Eq. \eqref{eq_generalized_NFW}, whose parameters are estimated by fitting the stacked satellite radial density profile measured in Sect. \ref{sec_measure_density_profile}. This measurement is possible only with an adequate number of subhalos, which requires a group mass $M_{200c}>M_1$. In addition, halos of mass lower than $M_1$ have $R_{200c}$ smaller than the minimum separation at which we measured the two-point correlation function in Sect. \ref{sec_measure_twop}. From a physical point of view this is equivalent to exclude the one-halo contribution for groups with a central halo only. \\
\indent For the halo mass function, $n(M,z)$, we adopt the functional form provided by \citet{Tinker08}. It can be integrated to compute the total halo number density,
\begin{equation}
    \label{eq_nh}
    n_{h}(z)=\int_{M_{\min }}^{M_{\max }} n(M, z) \langle N_h\rangle \mathrm{d}M, 
\end{equation}
in which $\langle N_h \rangle$ is expressed through Eq. \eqref{eq_occupation_number}. The satellite-satellite power spectrum is
\begin{equation}
    P_{ss}(k, z)=\frac{1}{n_h^2(z)} \int_{M_\mathrm{min}}^{M_{\max }}\langle N_s (N_s-1)\rangle n(M, z) \tilde{u}^2_s\left(k, M, z\right) \mathrm{d} M.
\end{equation}
Finally, the two-halo power spectrum is given by 
\begin{multline}
    \label{eq_Pk2halo}
    P_{2h}(k, z) = P_\mathrm{lin}(k, z) \Bigg[\frac{1}{n_h(z)} \int_{M_{\min}}^{M_{\max}} \langle N_{h}\rangle n(M, z) b(M, z) \\
    \times\tilde{u}_s(k, M, z) \, \mathrm{d}M \Bigg]^2 (A+Bk+Ck^2),
\end{multline}
where $b(M,z)$ is the linear bias provided by \citet{Tinker10}, while $A$, $B$ and $C$ form a scale-dependent nonlinear correction \citep[see for example][]{Smith07}, which reproduces the small-scale two-halo clustering of groups. Indeed, a more accurate description of the transition region where the one-halo and the two-halo contributions are almost equivalent, called quasi-linear regime, is traditionally more difficult to model \citep{Mead15}, and requires to consider nonlinear bias \citep{vandenBosch13, Bhowmick18} and exclusion effects \citep{Tinker05}. Discrete tracers, such as galaxies and dark matter halos, cannot overlap; thus the distance $r$ among them is always larger than the sum of their radii \citep{Baldauf13}. This reduces the two-halo correlation function at small scales. Moreover, the two-halo term itself has a small-scale behavior that cannot be reproduced by the assumption of the model bias presented in \citet{Tinker10}, as it contains nonlinearities that, if not taken into account, would alter the two-point correlation function at the typical one-halo separation, ending up with systematic errors on the HOD parameters. Moreover, the impact of exclusion effects is limited by the abundance of groups with mass between $10^8$ M$_\odot \,h^{-1}$ and $10^{10}$ M$_\odot \,h^{-1}$, which have $R_{200c}<0.05$ Mpc $h^{-1}$. This also allows us to count the two-halo pairs with spatial separations that are fully in the nonlinear regime. For this reason, for each redshift and cosmological model, we calibrate the nonlinear deviation from $b_h^2P_\mathrm{lin}(k,z)$, summarized in the parameters $A,B$ and $C$, by modeling the clustering of dark matter groups only, without considering the contribution of satellites. \\
\indent On large scales, the shape of the correlation function is determined by $\langle N_h\rangle$, and essentially differs from the matter-matter correlation for a linear bias factor only. On the other hand, on small scales, the one-halo term is sensitive to $\langle N_h(N_h-1)\rangle$ \citep{Berlind02}. A higher slope value, $\alpha$ or a lower normalization $M_1$, determines an increase in the number of satellites at high $M$. This largely boosts the number of satellite-satellite pairs in the one-halo term of the two-point correlation function. It also impacts the two-halo term, since higher masses, which are strongly biased, become more populated. In other words, for a given number of halos, we remove substructures from low-mass systems, redistributing them into more massive ones. Furthermore, higher values for $\gamma$ and $f_c$, as in Eq. \eqref{eq_generalized_NFW}, steepen the satellite density profile, increasing the amplitude and the slope of the one-halo correlation function. \\
\indent In Fig. \ref{fig_xi_model} we show the different contributions to the two-point correlation function as a function of redshift, for a CDM cosmology and $M>10^8$ M$_\odot \, h^{-1}$. In particular, on large scales, $\xi_{2h}(r)$ is largely suppressed by the finite size of the box. This condition is called ‘integral constraint' and requires to be forward-modeled in order to obtain unbiased results, as discussed in Sect. \ref{sec_measure_twop}. On the other hand, on small scales, the predictions of the linear model are not accurate enough. In particular, at high redshift, the scale-dependent terms, $B$ and $C$, are more relevant, while at $z=0$ the constant $A$ deviates more from 1. For example, at $z=2$ we found $A=0.80\pm 0.04$, $B=0.41\pm 0.02$ and $C=-0.010\pm 0.002$, while at $z=0$, $A=0.67\pm 0.03$, $B=0.07\pm0.01$ and $C=-0.004\pm 0.001$. Although $\xi_{1h}(r)$ always overcomes the linear $\xi_{2h}(r)$ at $r<0.2$ Mpc $h^{-1}$, it prevails on the nonlinear two-halo term only at $z=0$. It is dominated by satellite-satellite pairs, meaning that our dark matter groups are largely populated by subhalos. On the other hand, the correlation function of the central-satellite pairs, $\xi_{cs}$, increases with the redshift in absolute terms and relative to its satellite-satellite counterpart, $\xi_{ ss}$. Their ratio extends from a maximum of approximately $0.09$ at $z=0$ to $0.36$ at $z=2$. Given the same mass cut of $10^8$ M$_\odot \, h^{-1}$, moving toward high $z$, we find a larger contribution of isolated or poorly populated halos. Conversely, when we only select subhalos with a mass greater than $10^9$ M$_\odot \, h^{-1}$, the term $\xi_{cs}$ becomes progressively more important and dominates $\xi_{ss}$ for $M>10^{10}$ M$_\odot \, h^{-1}$.\\ 
\section{Results}
\label{sect_results}
In this section, we present the main results of our study on the halo occupation function, the subhalo distribution, and clustering. We discuss the outcome of the measurements in the CDM and alternative dark matter scenarios, and we then focus on our constraints on the HOD and density profile parameters. The Bayesian analysis was performed by adopting a Gaussian likelihood for each redshift bin
\begin{equation}
    \ln \mathcal{L}\propto-\frac{1}{2} \sum_i^{N_\mathrm{bin}}\left(\frac{m_i-d_i}{\sigma_i}\right)^2,
\end{equation}
where $m_i$ is the model on the radial scale $r_i$, $d_i$ is the measurement, and $\sigma_i$ is the associated error.
\subsection{Number counts}
\label{sec_number_counts}
\begin{figure*}[htbp]
\centering
\includegraphics[width=0.9\textwidth]{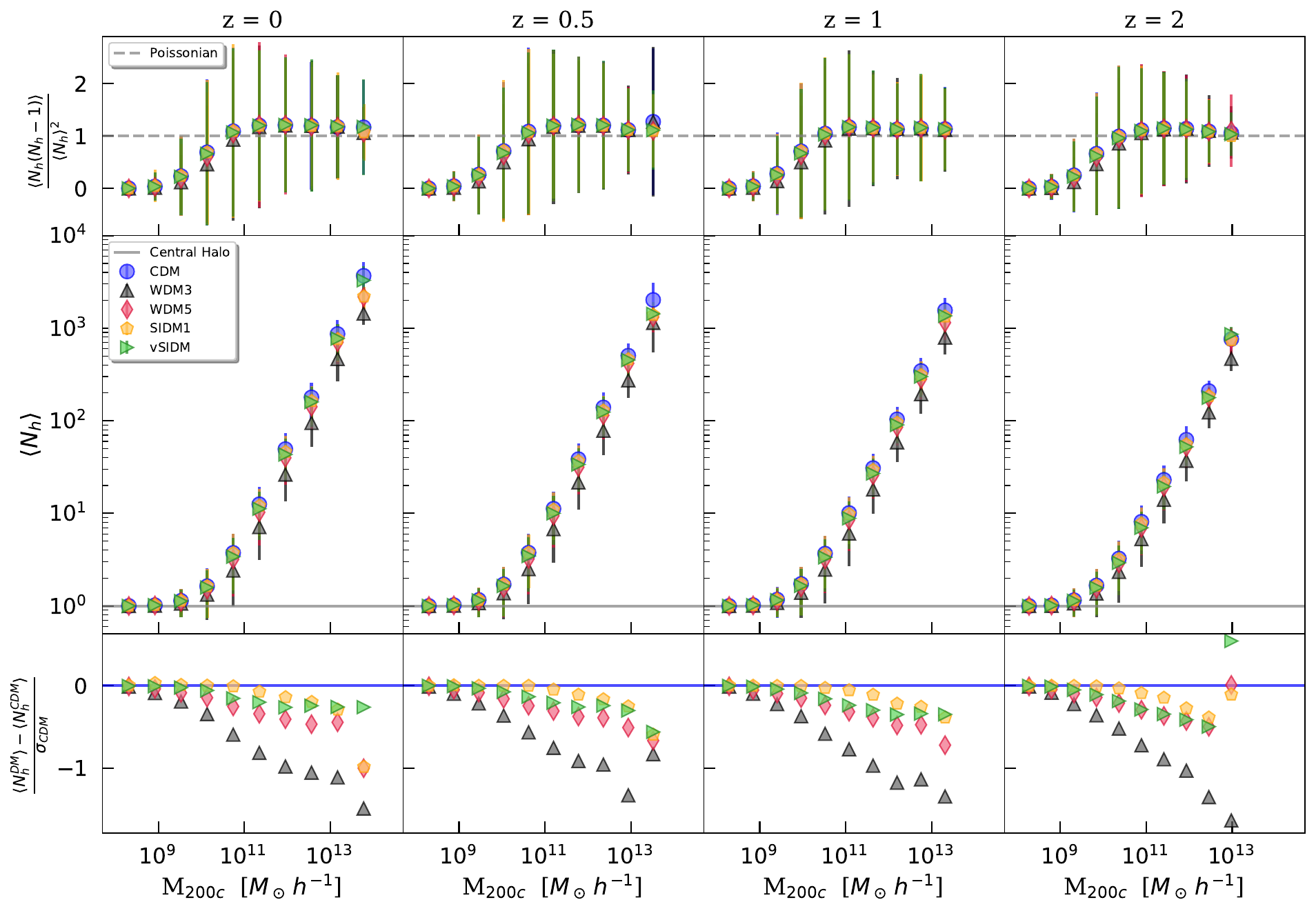}
\caption{From left to right: Average number of dark matter pairs (top panels), halo occupation function (middle panels) and residuals (bottom panels) with respect to the CDM reference level (solid blue line) as a function of $M_{200c}$ in box 50/A in the range $0<z<2$ for CDM (blue circles), WDM3 (black triangles), WDM5 (crimson diamonds), SIDM1 (gold pentagons), and vSIDM (green triangles). The corresponding errors are computed with the standard deviation of the measurements for each mass bin. The dashed horizontal gray line represents the expected ratio for a Poissonian distribution. The solid horizontal gray line marks the central halo that always exists for $M_{200c}>M_\mathrm{min}$. 
}
\label{fig_number_counts}
\end{figure*}
\begin{figure*}[htbp]
\centering
\includegraphics[width=0.9\textwidth]{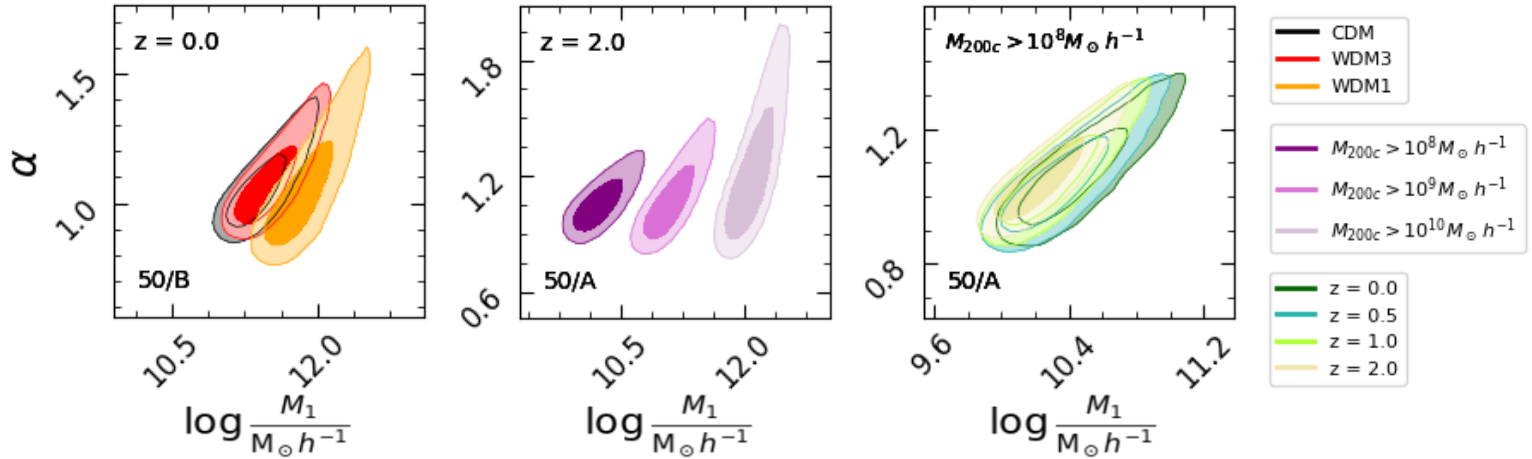}
\caption{Posterior distributions of the halo occupation function, $\langle N_h \rangle$, for different dark matter cosmologies (left panel), mass selections (middle panel), and redshifts (right panel). Different dark matter models are tested in box 50/B, with a mass selection of $M_\mathrm{min}>10^9$ M$_\odot \,h^{-1}$. The other two panels  consider the CDM cosmology in box 50/A.}
\label{fig_counts_contours}
\end{figure*}

For each \textsc{FoF} group in the box at redshift $z$, the \textsc{aida-tng} project gives access to the mass $M_{200c}$. Around each halo position, $(x_h, y_h, z_h)$, we build a sphere of comoving radius $R^\mathrm{com}_{200c}$ and count the number of subhalos within the region. Then, we split the group catalog by averaging the counts into ten logarithmic mass bins. In the same way, we also measured $\langle N_h(N_h-1)\rangle$. \\
\indent Fig. \ref{fig_number_counts} clearly separates the contribution of the central halo from that given by the associated substructures, namely the satellite halos. The applied mass selection ensures that there is at least one halo for $M_{200c}>M_\mathrm{min}$, while subhalos are counted from $M_{200c}>M_1$. In the top panels we show the average number of pairs, $\langle N_h(N_h-1)\rangle$. As explained in Sect. \ref{sec_hod_number_counts}, this quantity enables us to determine whether the subhalo population follows a Poissonian distribution. Although the differences between cosmologies are negligible, we observe that the hypothesis is fully satisfied when $M_{200c}>M_1$, where the number of satellites begins to deviate from zero. \\
\indent In Fig. \ref{fig_number_counts} we also show the occupation function $\langle N_h\rangle$ for each dark matter model. Alternatives to CDM are characterized by a reduction in the average number of satellites, with larger departures for the more massive systems. In WDM scenarios, this is particularly pronounced. For WDM3, the comparison with CDM shows residuals that exceed -0.5 for $M_{200c}>10^{11}$ M$_\odot \,h^{-1}$. A different mass selection, namely $M_\mathrm{min}>10^9$ M$_\odot \,h^{-1}$, significantly attenuates the effect, clarifying its strong dependence on the mass as a consequence of free streaming. For such a mass cut, only the warmest model, namely WDM1, exhibits a difference with respect to WDM3 and CDM, as we show in Fig. \ref{fig_counts_contours}. Here, we fit Eq. \eqref{eq_occupation_number} to quantify the dependence of $\alpha$ and $M_1$ on the redshift and mass selection of our sample; $M_0$, which represents only an offset in the mass of the group in the numerator, is not constrained and its presence does not significantly modify the mean values of $\alpha$ and $M_1$, although it widens their posterior distributions. For the sake of example, in Fig. \ref{fig_counts_contours} we show some results for boxes 50/A and 50/B. In particular, we note that for $M_\mathrm{min}>10^8$ M$_\odot \,h^{-1}$, we have $\log[{M_1}/({\mathrm{M}_\odot h^{-1}})]=10.42^{+0.25}_{-0.18}$ at $z=0$ and $\log[{M_1}/({\mathrm{M}_\odot h^{-1}})]=10.24^{+0.20}_{-0.16}$ at $z=2$, reflecting the fact that despite the presence of more massive structures, the average number of satellites in a given mass bin is slightly reduced in the local Universe by tidal disruption and merging phenomena. In fact, due to the hierarchical growth of cosmic structures, we first expect the formation of low-mass halos, which represent the progenitors of the systems that we count at $z=0$, which can host more than $10^3$ subhalos. In contrast, $\alpha=1.06^{+0.10}_{-0.08}$ at $z=0$ and $\alpha=1.05^{+0.10}_{-0.07}$ at $z=2$, which means that $\langle N_h \rangle$ increases approximately with the same slope at different redshifts. \\
\indent As also shown in Fig. \ref{fig_counts_contours}, a higher minimum mass tends to significantly increase the value of the normalization, while the slope rises only slightly. At $z=0$, for $M_\mathrm{min}>10^9$ M$_\odot \,h^{-1}$ we have $\log[{M_1}/({\mathrm{M}_\odot h^{-1}})]=11.47^{+0.23}_{-0.16}$ and $\alpha=1.17^{+0.14}_{-0.09}$, while for $M_\mathrm{min}>10^{10}$ M$_\odot \,h^{-1}$ we have $\log[{M_1}/({\mathrm{M}_\odot h^{-1}})]=12.25^{+0.20}_{-0.17}$ and $\alpha=1.16^{+0.17}_{-0.12}$, indicating that more massive halos tend to have relatively more massive substructures \citep{Zehavi05}. Finally, the broadening of the posterior distribution is a consequence of the lower statistics. \\
\indent Since in SIDM models the number density of dark matter halos does not differ significantly from the CDM case, as shown in \citet{Despali25}, the reduction in the average number of subhalos can be interpreted as a consequence of the more diffuse central density peak in self-interacting models, which results in satellite density profiles which are in general shallower around the core. In other words, moving toward outer regions, for a given mass, the cumulative radial count of subhalos presents a slope higher than the corresponding CDM model. However, these differences only result in a minimal increase and decrease in $M_1$ and $\alpha$, respectively, which, however are not sufficient to discriminate between dark matter models.
\subsection{Radial distribution of subhalos}
\label{sec_measure_density_profile}
\begin{figure*}[htbp]
\centering
\includegraphics[width=0.7\textwidth]{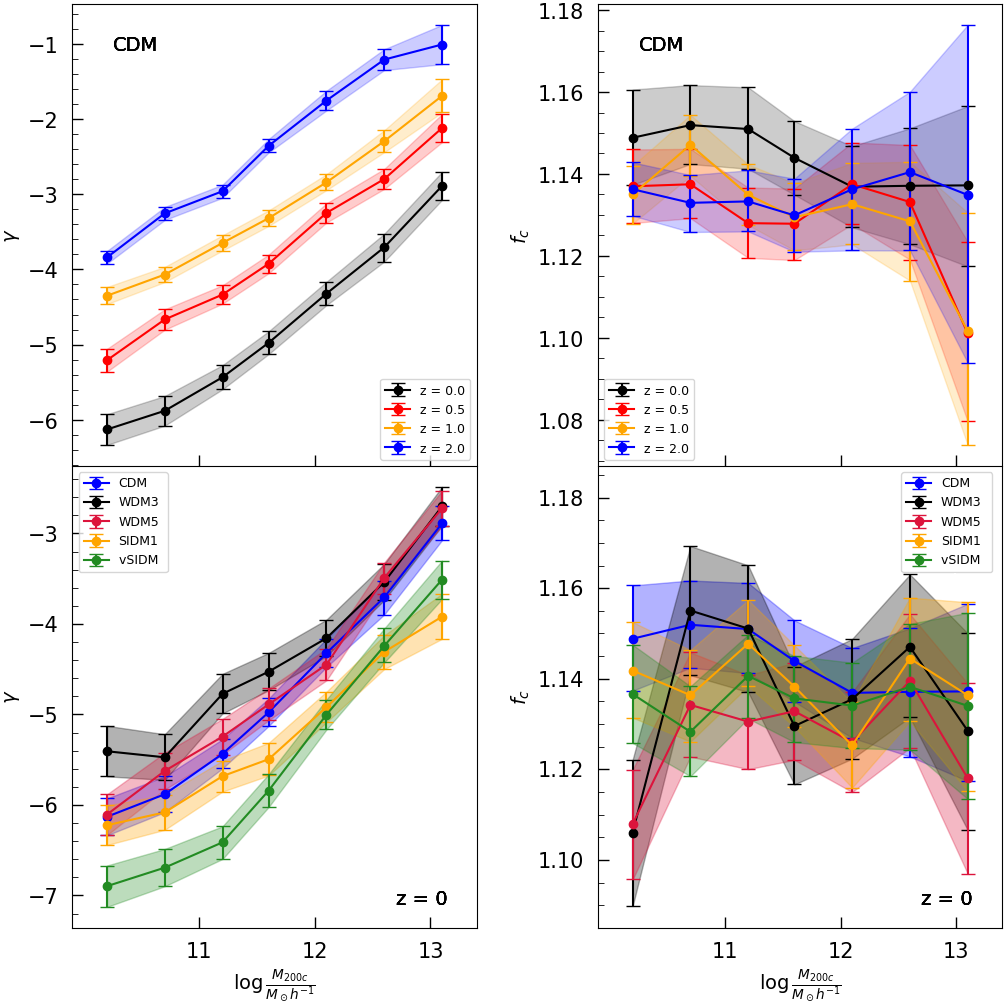}
\caption{Marginalized posterior distribution of the gNFW parameters, $\gamma$ (left) and $f_c$ (right), for different mass bins. Top panels: Evolution with redshift for a fixed $\Lambda$CDM cosmology. Bottom panels: Evolution in alternative dark matter models at $z=0$. The errors represent the 68\% percentiles.}
\label{fig_profile_parameters}
\end{figure*}

In order to disentangle the contribution of the central and satellites to the clustering signal of dark matter halos, we need to study how satellites are distributed within their hosts. To do this, we split the halos into ten logarithmic-spaced mass bins, with a lower limit given by the transition mass from the central-dominated to the satellite-dominated regime of Fig. \ref{fig_number_counts}, at $z=0$, which depends on the resolution of the boxes, and corresponds to $10^{10}$ M$_\odot \,h^{-1}$ for 50/A. For each mass bin, we consider $N_\mathrm{bin}=8$ logarithmic-spaced radial bins. In practice, from the middle value of each mass bin we estimate the corresponding comoving radius $R^\mathrm{com}_{200c}$, which is adopted as $R_\mathrm{max}$, while $R_\mathrm{min}=R_\mathrm{max}/(N_\mathrm{bin}-1)$, to exclude the innermost regions, where there is a lack of satellites due to mass selection, tidal disruption, and resolution effects. The counts are then averaged for every mass bin, and normalized according to the comoving volume of the $i-$th radial shell, $V_i=\frac{4}{3}\pi(R_{\mathrm{max},i}^3-R_{\mathrm{min},i}^3)$, obtaining $\tilde{n}_s(r)$, our stacked profile.\\ 
\indent From measurements of the radial subhalo distribution, we can quantify possible deviations from a pure NFW model. Dealing with discrete tracers rather than a continuous dark matter density field, we compute the characteristic number density of satellites as
\begin{equation}
\label{eq_generalized_NFW_discrete}
    n_s(r)= \frac{\tilde{N}_s}{M_{200c}}\rho(r),
\end{equation}  
where $M_{200c}$ corresponds to the average in a given mass bin, $\rho(r)$ is given by Eq. \eqref{eq_generalized_NFW}, and 
\begin{equation}
    \tilde{N}_s = 4 \pi \int_{R_\mathrm{min}}^{R_\mathrm{max}} r^2 \tilde{n}_s(r) \mathrm{d}r
\end{equation}
is the normalization factor measured from the radial density profile, $\tilde{n}_s(r)$.
\begin{figure*}[htbp]
\centering
\includegraphics[width=0.9\textwidth]{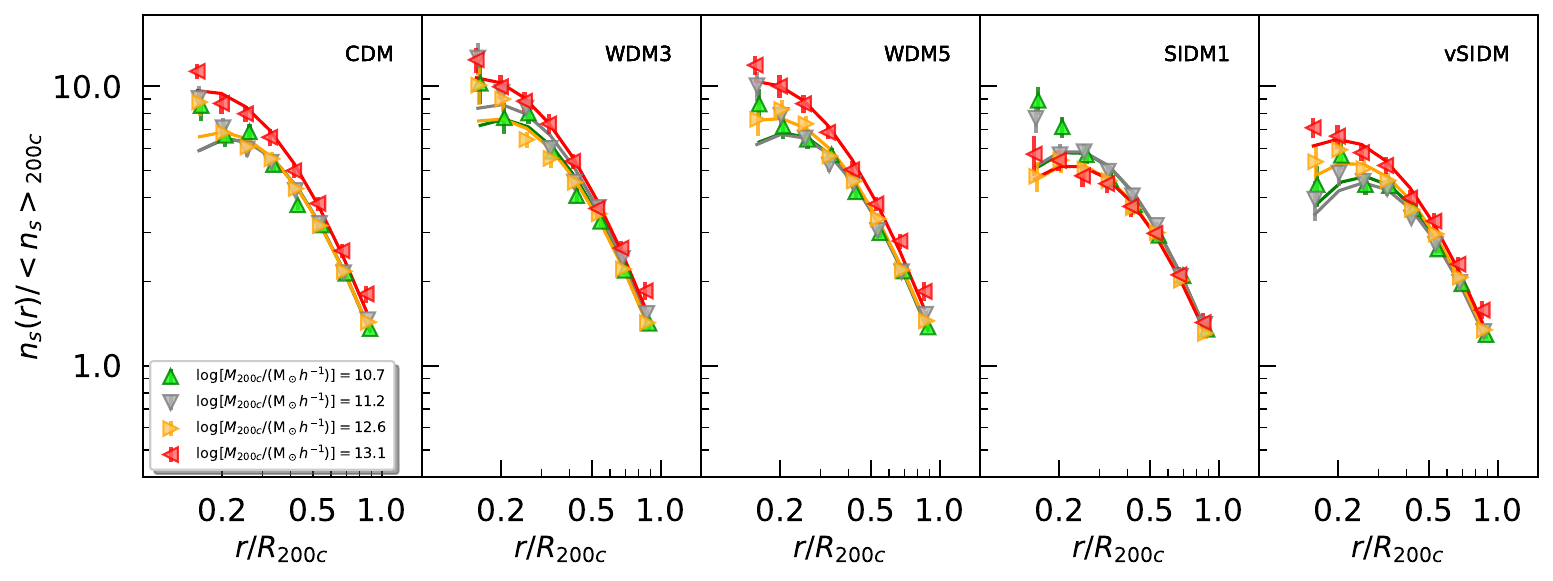}
\caption{From left to right: Radial distribution of dark matter subhalos at $z=0$ for various mass bins and dark matter models. The triangles represent bins with a group mass of $\log[{M_{200c}}/({\mathrm{M}_\odot h^{-1}})]=10.7$ (green), $11.2$ (gray), $12.6$ (gold), and $13.1$ (red), respectively. The solid, colored lines reflect the gNFW best-fit model. The corresponding errors are computed with the standard deviation of the measurements, for each mass bin.}
\label{fig_stacked_density}
\end{figure*}
Figure \ref{fig_stacked_density} shows the radial distribution of subhalos normalized to the average value of the satellite density at a radius of $R_{200c}$, for standard $\Lambda$CDM and the alternative dark matter models. Variations among different masses are highlighted, while for $r>0.5\times R_{200c}$ the rescaled radial profiles are relatively self-similar; a feature that, however, is lost at higher redshifts because of the selection effects, which impact more when the average mass of structures is lower. In general, more massive structures are denser in terms of the number of satellites. This is because the effect of dynamical friction is progressively more relevant for massive subhalos hosted by large dark matter groups. As a consequence, they sink toward the halo center, reducing the radius of their orbits and enhancing the inner part of the density profile, while the outer region is less affected \citep{Han16, Chang18}. This fact incidentally explains why luminous galaxies are poor tracers of the NFW dark matter density distribution, while they match the corresponding subhalo profiles found in cosmological simulations \citep{Watson12, Chang18}. Conversely, since the halo mass is inversely correlated with the age, smaller hosts at $z=0$ are, on average, older and have had more time to destroy subhalos located near the center. Interestingly, the situation is reversed in SIDM1 model, for which low-mass groups, with $\log[{M_{200c}}/({\mathrm{M}_\odot h^{-1}})]$ of $10.7$ and $11.2$ are centrally more peaked than the massive ones, with $\log[{M_{200c}}/({\mathrm{M}_\odot h^{-1}})]$ of $12.6$ and $13.1$, which have thermalized earlier. In fact, self-interaction between dark matter particles produces more diffuse central peaks and more extended cores for high-mass halos \citep{Despali25}. While in vSIDM scenario the cross-section is a decreasing function of the relative velocity of particles, in SIDM1 it is constant. Therefore, moving toward the typical galaxy cluster size, this scenario is strongly constrained \citep{meneghetti01,eckert22,shen22} since it could cause major departures from the $\Lambda$CDM cosmology; these are, however, ruled out by observations that found cusps at the center of galaxy clusters \citep{Newman13}. \\
\indent In Fig. \ref{fig_stacked_density} we show the best fit, calculated according to Eq. \eqref{eq_generalized_NFW_discrete}. In addition to gNFW, we also test the density profile model presented in \citet{Einasto65}.
Although it performs well if we analyze the full density profile up to scales larger than $R_{200c}$, it fails at small separations and for low mass halos, for which the friction force is not enough to flatten the spatial distribution of satellites within the core. We estimate the parameters of the gNFW profile, $\gamma$ and $f_c$, choosing large, flat priors, between $[-10, 2]$ and $[0.1,1.5]$, respectively. 
In Fig. \ref{fig_profile_parameters}, for a $\Lambda$CDM cosmology, we show the mass and redshift evolution of the density profile parameters, considering only mass bins where halos are present at all redshifts and for all cosmologies. In particular, $\gamma$ assumes negative values, reflecting the well-established results that the halo distribution is antibiased \citep{Ghigna98, Springel08}, namely it is more cored toward the center than the matter density. The analytical model presented in \citet{Han16} explains the spatial distribution of subhalos as a result of tidal stripping, generated from dynamical friction which, conversely, does not affect the dynamics of dark matter particles. Furthermore, $\gamma$ increases with the mass of the group $M_{200c}$, which means that less massive systems generally present a shallower radial distribution of satellites. However, we should consider that a variation in $M_{200c}$ from approximately $10^{13}$ M$_\odot \,h^{-1}$ to $10^{10}$ M$_\odot \,h^{-1}$ translates into a reduction of $R_{200c}$ of about one order of magnitude; therefore we explore a shorter range of spatial separations in low-mass groups. In this case, the value of $\gamma$ reflects the lack of subhalos toward more central regions, due to stripping phenomena, box resolution and selection effects. Also for this reason, the majority of subhalos are located in the outer regions of the group, which correspond to shells of larger volume. \\
\indent In high-$z$ Universe, dark matter subhalos form in an environment with a higher mean matter density, which allows a more efficient accretion process. In addiction, they are less massive and more concentrated. These features translate into a reduction of the tidal stripping. Furthermore, over time, their average number decreases as they merge. Having measured the density in comoving radial shells, only the net effect of satellite creation and destruction is considered, while the total mass of the central peak increases toward $z=0$. Thus, in time the satellite density profiles become flatter. This explains the redshift evolution of the $\gamma-M_{200c}$ relation that we see in Fig. \ref{fig_profile_parameters}. On the other hand, the concentration factor, $f_c$, is roughly constant for halos in a wide range of masses and redshift, which share the same deviation from the adopted $c-M_{200c}$ relation. This simply describes a general shrinking in the characteristic radius $r_s$ with respect to that of the dark matter density profile, reflecting the loss of orbital energy of the subhalos that experience dynamical friction \citep{Chang18}.\\ 
\indent In Fig. \ref{fig_profile_parameters} we also see the variation of the gNFW density profile parameters in the different dark matter scenarios. Although the general mass and redshift tendencies are kept, there is no clear dependence of $f_c$ on cosmology. Instead, WDM models prefer a higher $\gamma$, which means that the subhalos radial distribution is more concentrated and steeper around the center, as also found in the analysis of the Aquarius-WDM runs \citep{Lovell14}, the \textsc{coco} simulations \citep{Bose17} and zoom-in simulations of early type galaxies from the \textsc{eagle} box \citep{Despali20}. Indeed, the free-streaming of dark matter particles determines an overall delay in the structure formation process, especially for less massive systems \citep{Lovell24}. Therefore, WDM subhalos of a given mass generally form later than their CDM counterparts, when the average density of the Universe is smaller, allowing more concentrated structures to grow. Conversely, in SIDM scenarios, $\gamma$ is generally lower, which means that the innermost satellite density profile is shallower and less concentrated. This result has been already found in simulations by \citet{Fischer22, Fischer24}, as a reduction in the number of subhalos becoming progressively stronger at smaller distances to the host, and can be interpreted as a consequence of core thermalization due to particle interaction. This is in fact expected to produce a flatter underlying dark matter distribution and make satellites more sensitive to tidal disruption. For this reason, the relative number of subhalos is distributed toward the external radial shells of the \textsc{FoF} group. At $z=0$, for $M_{200c}>10^{12.5}$ M$_\odot \,h^{-1}$, the radial distribution of the SIDM1 subhalos becomes shallower than that of the vSIDM. This once again highlights the tendency of SIDM1 to form cores even in the most massive systems, while vSIDM interactions are only significant for low-mass bound structures. The behavior of WDM and SIDM cosmologies, quantitatively appreciable already in Fig. \ref{fig_stacked_density}, is also reflected in their clustering properties, as explained in Sect. \ref{sec_measure_twop}.
\subsection{The two-point correlation function}
\label{sec_measure_twop}
\begin{figure*}[htbp!]
\centering
\includegraphics[width=0.9\textwidth]{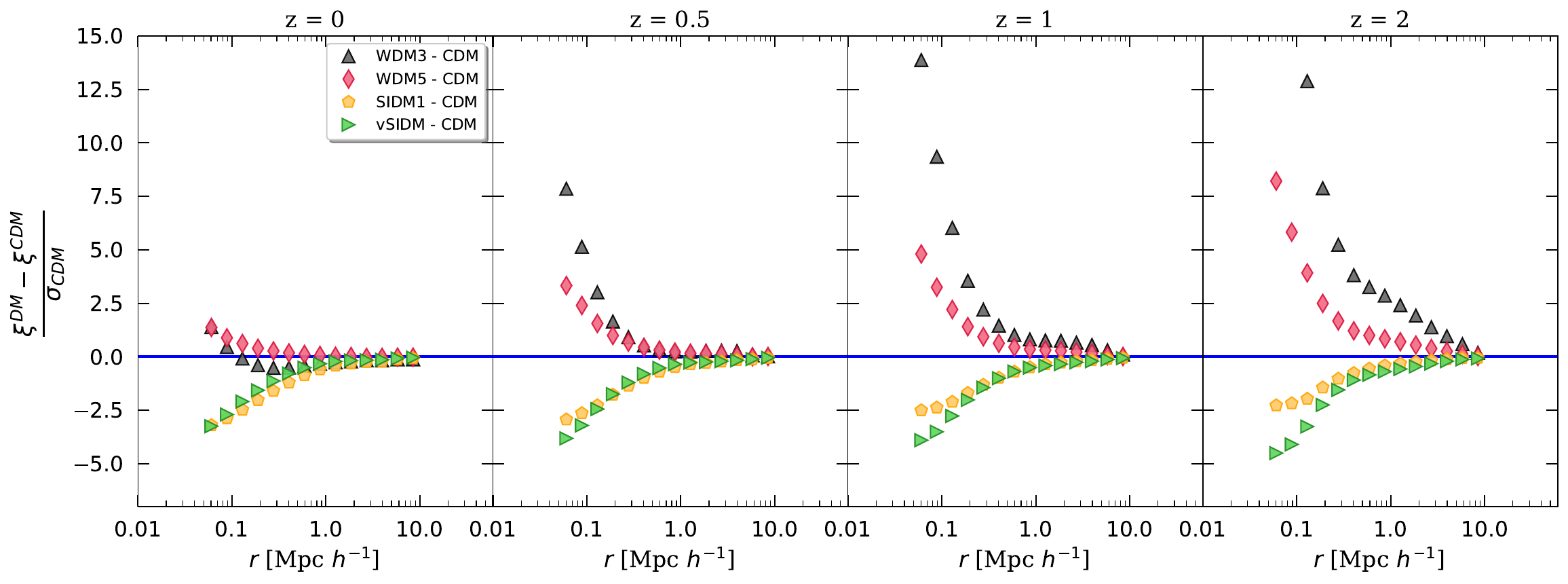}
\caption{From left to right: Residuals of the two-point correlation function for alternative dark matter models for redshifts from $z=0$ to $z=2$. The blue, horizontal line represents the reference level for $\Lambda$CDM model. WDM3 (black triangles), WDM5 (crimson diamonds), SIDM1 (gold pentagons) and vSIDM (green triangles) are also shown.}
\label{fig_clustering}
\end{figure*}

The clustering information of a sample of discrete tracers is encoded in the two-point correlation function, $\xi
(r)$. According to the definition, for a pair of elements located in volume elements $\delta V_1$ and $\delta V_2$, at a given separation $r$, 
it is expressed as an excess probability with respect to a uniform distribution,
\begin{equation}
\label{eq_dP_12}
\delta P_{12}(r) = n_V^2\,[1+\xi(r)]\,\delta V_1 \delta V_2\,,
\end{equation}
where $n_V$ is the mean tracer number density in a unit of comoving volume. \\
\indent We build the random catalog by uniformly extracting the comoving coordinates of dark matter halos in a box. Since for each redshift the number of tracers is higher than $4\times 10^5$, the shot noise is limited, and the size of the random catalog is kept equal to the size of the original one, in order to speed up the computation. We measure the real-space two-point correlation function with the \citet{Landy93} estimator,
\begin{equation}
\label{eq_LandySzalay}
    \xi_\mathrm{obs}(r)=\frac{DD(r)+RR(r)-2DR(r)}{RR(r)},
\end{equation}
which provides an unbiased measurement of $\xi_\mathrm{obs}(r)$ once the normalized number of data-data, random-random, and data-random pairs, $DD(r)$, $RR(r)$ and $DR(r)$ respectively, is counted. The errors are estimated using the jackknife method, considering a number of regions equal to 125 \citep{Norberg09}. We adopt appropriate mass cuts to ensure completeness, corresponding to a lower limit of $10^8$ M$_\odot h^{-1}$ for 50/A. The two-point correlation is measured into fifteen logarithmic-spaced bins between $0.05$ Mpc$\,h^{-1}$ and $15$ Mpc $h^{-1}$. The lower limit is motivated by the resolution of the simulation, whereas the upper limit is due to the finite size of the box and corresponds approximately to the scale at which the correlation function becomes negative. At these scales nonlinear effects are not negligible and the halo bias depends on $r$, as well as on redshift and mass. A higher-mass cut selects only more massive halos, which are more biased, and therefore provides a stronger clustering signal. Furthermore, with cosmic time, the one-halo term becomes more pronounced, implying a progressive increase of nonlinearities. \\
\indent The use of cosmological simulations with size much smaller than the volume of the Universe has non-negligible consequences on the measurement of the clustering signal. 
From a theoretical point of view, a scale-free power spectrum satisfies the condition $P(k=0) = 0$, which corresponds to imposing homogeneity of matter fluctuations on sufficiently large scales, where the cosmological principle is satisfied \citep{Riquelme23}. According to the definition of probability, the integral of Eq. \eqref{eq_dP_12} over the full volume of the Universe is equal to 1. The same condition also holds if we consider the finite volume of the box, $V_b$. This implies that 
\begin{equation}
    \label{eq_xi_IC}
    \int \xi_\mathrm{obs}(r) \mathrm{d}V_b = 0,
\end{equation}
which means that we should find a scale at which $\xi_\mathrm{obs}(r)$ becomes negative, which is lower than the corresponding scale for the model of $\xi(r)$. In other words, the evaluation of the correlation function in a limited volume determines a reduction in the clustering signal. The observed correlation function is linked to the theoretical one through a window function, $W(\vec{r})$. In Eq. \eqref{eq_xi_IC}, in order to avoid volume integration in surveys with complicated geometries, it is common to correct the model prediction by following the radial evolution of random-random pairs. This information is contained in the shape of the random catalog. Such `integral constraint' condition therefore takes the form \citep{Ross13, Riquelme23} 
\begin{equation}
    \xi_{IC}(r) = \xi(r)-\frac{\sum_i \xi(r_i)RR(r_i)}{\sum_i RR(r_i)},
\end{equation}
and spans over the range at which the pairs are counted. \\
\indent In Fig. \ref{fig_clustering}, we compare the measures obtained in box 50/A within the $\Lambda$CDM framework, with the results in alternative dark matter scenarios, finding that the main difference between models is located at the one-halo scales, corresponding to $r\lesssim 1$ Mpc $h^{-1}$. In fact, SIDM1 and vSIDM are systematically less clustered, and the difference to CDM is only slightly affected by time evolution, since they residuals remain between $-5$ and $-2.5$ for all the redshifts considered. In particular, it is noteworthy that SIDM1  is closer to CDM at high $z$, indicating a late-time thermalization of the core; conversely, the vSIDM model deviates more at $z=2$. In other words, vSIDM is more effective at high $z$, since the cross-section is higher for low-mass halos, where the velocity dispersion of the particles is smaller. On the other hand, in WDM cosmologies, the average delay in the structure formation process makes clustering differences more visible at high $z$, with residuals larger than $2.5$ for $r\lesssim 1$ Mpc $h^{-1}$ and $r\lesssim 0.2$ Mpc $h^{-1}$, for WDM3 and WDM5, respectively. These results reinforce what we already found from the analysis of Sect. \ref{sec_measure_density_profile}, according to which WDM generally produces a more peaked subhalo density distribution. As we have already seen in Sect. \ref{sect_model_twop} the satellite-satellite pairs dominate the one-halo range. Consequently, the radial distribution of the satellites directly affects the clustering properties of the full halo population. However, the excess of clustering is also found in the two-halo term for the warmest models. Again, we can interpret this trend as a consequence of free-streaming of the particles. Indeed, the suppression of low-mass dark matter halos has a direct impact on the halo mass function, which operates as a weighting factor for the halo bias, resulting in practice in a larger effective bias of our sample.\\
\indent A proper modeling of the observed two-point correlation function is well beyond the linear theory, requiring a generalization of the density profiles and the introduction of the HOD formalism. Our ultimate goal is to explore the degeneracy of HOD parameters in different cosmological frameworks. For this reason, as already described in Sect. \ref{sect_model_twop}, we calibrate the nonlinear deviation from $b_h^2P_\mathrm{lin}(k,z)$, by modeling the clustering of dark matter groups only, without considering the contribution of satellites. In this way, we quantify the $A+Bk+Ck^2$ factor of Eq. \eqref{eq_Pk2halo}, which incorporates the exclusion effects and the inaccuracies of the nonlinear bias model for each redshift and dark matter scenario, adopting their median value for the HOD characterization of the full halo populations. \\
\indent Then we fit Eq. \eqref{eq_xi_tot} to constrain the parameters $M_1$ and $\alpha$, whose posterior distributions are not affected by the presence of $M_0$. We set the latter to zero, to reduce the number of parameters and thus the complexity of the model. In order to assume a physically motivated normalized density profile, $\tilde{u}_s\left(k, M, z\right)$, we adopt flat priors on the parameters $f_c$ and $\gamma$, according to the range permitted by the posterior distributions presented in Sect. \ref{sec_measure_density_profile}, while $\log[{M_1}/({\mathrm{M}_\odot h^{-1}})]$ and $\alpha$ are left free to vary in the range $[9,13]$ and $[0.7,1.6]$, respectively. For the WDM models, we use Eq. \eqref{eq_transfer_function} to suppress the linear power spectrum, $P_\mathrm{lin}(k,z)$, at small scales, while for the halo mass function, $n(M)$, we apply Eq. \eqref{eq_suppression_hmf}. This reduces the low-mass halo population, resulting in a higher effective bias, which determines an enhancement of the two-halo term. SIDM models, on the other hand, share the same power spectrum and mass function as $\Lambda$CDM.\\
\begin{figure*}[htbp!]
\centering
\includegraphics[width=0.95\textwidth]{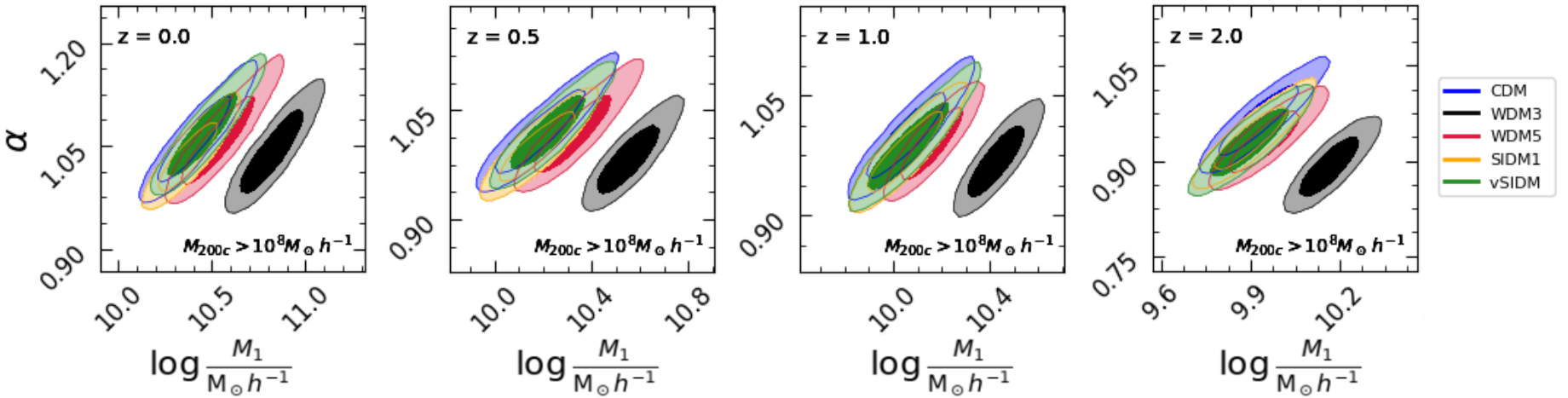}
\caption{Marginalized posterior distributions of the HOD parameters in the $M_1-\alpha$ plane for box 50/A and $M>10^8$ M$_\odot h^{-1}$. The 68\% and 95\% contours are represented for CDM (blue), WDM3 (black), WDM5 (crimson), SIDM1 (gold) and vSIDM (green) cosmological models.}
\label{fig_mcmc_clustering}
\end{figure*}
\indent In Fig. \ref{fig_mcmc_clustering} we show the marginalized posterior distribution in the $M_1-\alpha$ plane. The constraints on HOD parameters are in agreement with the results obtained in Sect. \ref{sec_number_counts} with halo occupation function. In particular, for the $\Lambda$CDM cosmology, we find $\log[{M_1}/({\mathrm{M}_\odot h^{-1}})]=10.41^{+0.15}_{-0.12}$ and $\alpha=1.06^{+0.05}_{-0.04}$ at $z=0$; $\log[{M_1}/({\mathrm{M}_\odot h^{-1}})]=9.93^{+0.09}_{-0.09}$ and $\alpha=0.97^{+0.04}_{-0.03}$ at $z=2$. However, they are also more competitive. In fact, the two-halo term of the correlation function relies on $\langle N_h\rangle$, as in the case of number counts, while the one-halo term depends on the number of pairs $\langle N_h(N_h-1)\rangle$. In addition to second-order statistics, the clustering model also includes stringent prior information from the parameters of the gNFW density profile, $f_c$ and $\gamma$. This emphasizes the necessary synergy between these different probes for future HOD-based studies.\\
\indent Clustering of halos reveals its full potential when applied to alternative dark matter scenarios. The $M_1$ parameter indeed increases in the WDM and SIDM models, alongside a slight decrease in $\alpha$. This reflects the fact that the average reduction in the number of satellites due to free streaming or core thermalization means that higher group masses are required for finding the same number of subhalos. This tendency was already observed in Sect. \ref{sec_number_counts}, where we analyzed the occupation functions; however, the ability to distinguish between different models was limited to box 50/B, for which only WDM1 presented a $1\sigma$ difference with CDM, for the parameter $M_1$. From this point of view, the study of the two-point correlation function represents a powerful probe, especially if we analyze the clustering of dark matter halos at high $z$. In particular, for the thermal WDM3 model we find $\log[{M_1}/({\mathrm{M}_\odot h^{-1}})]=10.83^{+0.11}_{-0.11}$ and $\alpha=1.04^{+0.04}_{-0.04}$ at $z=0$; $\log[{M_1}/({\mathrm{M}_\odot h^{-1}})]=10.16^{+0.07}_{-0.06}$ and $\alpha=0.89^{+0.03}_{-0.03}$ at $z=2$. In other words, the normalization in the number of satellites exhibits a difference greater than $2\sigma$ with respect to CDM, while at $z=2$, the one on the slope is at the level of $1\sigma$. WDM5 occupies an intermediate position, with $\log[{M_1}/({\mathrm{M}_\odot h^{-1}})]=10.53^{+0.13}_{-0.11}$ and $\alpha=1.06^{+0.04}_{-0.04}$ at $z=0$; $\log[{M_1}/({\mathrm{M}_\odot h^{-1}})]=9.95^{+0.08}_{-0.08}$ and $\alpha=0.93^{+0.03}_{-0.03}$ at $z=2$, and therefore requires more stringent mass cuts to achieve the same level of disagreement. Inspecting Fig. \ref{fig_mcmc_clustering}, it is also clear that fixed and velocity-dependent SIDM models are indistinguishable from each other. As the latter acts mainly where the velocity dispersion of the particles is lower, altering the structure and distribution of small-mass halos, we conclude that simulations with higher mass resolution are required to explore finer spatial scales in detail and to better highlight the differences between alternative dark matter models. In the future, we plan to exploit the full-physics \textsc{aida-tng} simulations, for a subsequent work, which will take into account the fact that not all dark matter halos and subhalos are able to host galaxies, as well as the complicated relationship between the mass and the luminosity of galaxies, directly linked to their formation processes. In this way, with more complete HOD models, we aim at disentangling the effects of alternative dark matter from baryonic physics, in preparation for a full comparison with observational data from galaxy clustering.
\section{Conclusions}
\label{sect_conclusions}
We studied the effect of warm and self-interacting dark matter models on the abundance, the radial distribution, and the clustering properties of subhalos. For this purpose, we used a set of dark matter-only cosmological boxes with a variable size and resolution that were developed within the \textsc{aida-tng} project. We applied the HOD formalism, which allowed us to separate the contributions of distinct populations of central and satellite halos, in order to model their occupation function and clustering. In particular, we counted the average abundances of subhalos within the comoving radius $R^\mathrm{com}_{200c}$ for different mass bins and found that the HOD parameter $M_1$ is higher at low redshift as a consequence of the impact of tidal stripping and merging phenomena on the average number of satellites. Alternative dark matter cosmologies predict a reduction in the number of subhalos within $R^\mathrm{com}_{200c}$. In WDM models, particle free streaming indeed suppresses the formation of low-mass halos, while in the SIDM case, the core thermalization redistributes satellites in the outer region of dark matter groups. However, while for box 50/B for $M>10^9$ M$_\odot h^{-1}$, CDM and WDM1 cosmologies exhibit an appreciable difference, in box 50/A for $M>10^8$ M$_\odot h^{-1}$, the occupation function does not provide a significant way to distinguish between models, requiring higher statistics and/or mass resolution. Moreover, with the applied mass selections, we verified that for the group mass $M_{200c}>M_1$, where the average number of satellites is larger than one, they follow a Poissonian distribution for all the models considered. \\
\indent We also evaluated the radial distribution of subhalos around the centers of the groups. To do this, we counted the average number density of satellites in spherical shells up to a radius equal to $R^\mathrm{com}_{200c}$. For each mass bin and dark matter model, we fit the stacked radial density with a gNFW density profile. In this way, we were able to constrain the evolution of the gNFW parameters $\gamma$ and $f_c$ with $M_{200c}$ and $z$. Although the latter is approximately constant and larger than one, as a consequence of dynamical friction, we found that the subhalo distribution becomes steeper for more massive systems at high-$z$, while it remains antibiased compared to the matter distribution. Furthermore, SIDM models generally prefer a shallower satellite radial profile as a result of the core thermalization in bounded dark matter halos, while WDM cosmologies on average admit a more cusped distribution toward the center of the groups as a result of the delayed structure formation process. This has direct consequences on the clustering properties of dark matter halos because it mainly affects the one-halo term of the two-point correlation function, where the differences between cold and alternative dark matter models are more relevant. In particular, in WDM cosmologies, the peaked subhalo distribution determines a clustering excess at $r\lesssim 0.2$ Mpc $h^{-1}$, with an increasing effect toward high $z$. At $z=2$, it indeed also extends to the two-halo scales; the free streaming of dark matter particles determines a low-mass suppression in the halo mass function, which increases the effective bias of our sample. Conversely, SIDM models are less clustered at small scales, and their residuals with respect to CDM are only slightly affected by the redshift evolution. The HOD modeling of the two-point correlation function provided constraints on $M_1$ and $\alpha$ that agree with the prediction of the halo occupation functions. In addition, it produced more stringent results, with a $2\sigma$ difference on $M_1$ between CDM and WDM3 from $z=0$ to $z=2$. This revealed the effectiveness of this cosmological probe in distinguishing between possible alternative dark matter scenarios.\\
\indent In the future, we plan to conduct a study on full-physics simulations to explore the degeneracy between the properties of dark matter particles and the physics of baryons in macroscopic observables, such as abundances, density profiles, and clustering. Variations in the physics of dark matter modify the abundance and the spatial distribution of halos, with a possible direct and macroscopic impact on luminous tracers, such as galaxies. Therefore, the completeness, spatial resolution, and depth of forthcoming photometric and spectroscopic surveys, combined with new state-of-the-art simulations, will offer unique possibilities to shed light on the real nature of dark matter. 

\begin{acknowledgements}
     We thank Ravi Sheth for useful discussions and hints on the paper. We would like to thank the anonymous referee for the constructive comments, which have improved the quality of the manuscript. We acknowledge the financial contribution from the grant PRIN-MUR 2022 20227RNLY3 “The concordance cosmological model: stress-tests with galaxy clusters” supported by Next Generation EU and from the grants ASI n.2018-23-HH.0 and n. 2024-10-HH.0 “Attività scientifiche per la missione Euclid – fase E”, and the use of computational resources from the parallel computing cluster of the Open Physics Hub (\url{https://site.unibo.it/openphysicshub/en}) at the Physics and Astronomy Department in Bologna. GD acknowledges the funding by the European Union - NextGenerationEU, in the framework of the HPC project – “National Centre for HPC, Big Data and Quantum Computing” (PNRR - M4C2 - I1.4 - CN00000013 – CUP J33C22001170001). We acknowledge ISCRA and ICSC for awarding this project access to the LEONARDO supercomputer. The \textsc{aida-tng} simulations where run on the LUMI supercomputer (Finland), thanks to the award of computing time through a EuroHPC Extreme Scale Access call.
\end{acknowledgements}

\bibliographystyle{aa} 
\bibliography{bibliografia} 

@ARTICLE{Lesci22,
       author = {{Lesci}, G.~F. and {Nanni}, L. and {Marulli}, F. and {Moscardini}, L. and {Veropalumbo}, A. and {Maturi}, M. and {Sereno}, M. and {Radovich}, M. and {Bellagamba}, F. and {Roncarelli}, M. and {Bardelli}, S. and {Castignani}, G. and {Covone}, G. and {Giocoli}, C. and {Ingoglia}, L. and {Puddu}, E.},
        title = "{AMICO galaxy clusters in KiDS-DR3: Constraints on cosmological parameters and on the normalisation of the mass-richness relation from clustering}",
      journal = {\aap},
         year = 2022,
        month = sep,
       volume = {665},
          eid = {A100},
        pages = {A100},
          doi = {10.1051/0004-6361/202243538},
archivePrefix = {arXiv},
       eprint = {2203.07398},
 primaryClass = {astro-ph.CO},
       adsurl = {https://ui.adsabs.harvard.edu/abs/2022A&A...665A.100L}
}

@ARTICLE{Landy93,
       author = {{Landy}, Stephen D. and {Szalay}, Alexander S.},
        title = "{Bias and Variance of Angular Correlation Functions}",
      journal = {\apj},
         year = 1993,
        month = jul,
       volume = {412},
        pages = {64},
          doi = {10.1086/172900},
       adsurl = {https://ui.adsabs.harvard.edu/abs/1993ApJ...412...64L}
}

@ARTICLE{MarulliCBL,
       author = {{Marulli}, F. and {Veropalumbo}, A. and {Moresco}, M.},
        title = "{CosmoBolognaLib: C++ libraries for cosmological calculations}",
      journal = {Astronomy and Computing},
         year = 2016,
        month = jan,
       volume = {14},
        pages = {35-42},
          doi = {10.1016/j.ascom.2016.01.005},
archivePrefix = {arXiv},
       eprint = {1511.00012},
 primaryClass = {astro-ph.CO},
       adsurl = {https://ui.adsabs.harvard.edu/abs/2016A&C....14...35M}
}

@ARTICLE{Tinker08,
       author = {{Tinker}, Jeremy and {Kravtsov}, Andrey V. and {Klypin}, Anatoly and {Abazajian}, Kevork and {Warren}, Michael and {Yepes}, Gustavo and {Gottl{\"o}ber}, Stefan and {Holz}, Daniel E.},
        title = "{Toward a Halo Mass Function for Precision Cosmology: The Limits of Universality}",
      journal = {\apj},
         year = 2008,
        month = dec,
       volume = {688},
       number = {2},
        pages = {709-728},
          doi = {10.1086/591439},
archivePrefix = {arXiv},
       eprint = {0803.2706},
 primaryClass = {astro-ph},
       adsurl = {https://ui.adsabs.harvard.edu/abs/2008ApJ...688..709T}
}

@ARTICLE{Tinker10,
       author = {{Tinker}, Jeremy L. and {Robertson}, Brant E. and {Kravtsov}, Andrey V. and {Klypin}, Anatoly and {Warren}, Michael S. and {Yepes}, Gustavo and {Gottl{\"o}ber}, Stefan},
        title = "{The Large-scale Bias of Dark Matter Halos: Numerical Calibration and Model Tests}",
      journal = {\apj},
         year = 2010,
        month = dec,
       volume = {724},
       number = {2},
        pages = {878-886},
          doi = {10.1088/0004-637X/724/2/878},
archivePrefix = {arXiv},
       eprint = {1001.3162},
 primaryClass = {astro-ph.CO},
       adsurl = {https://ui.adsabs.harvard.edu/abs/2010ApJ...724..878T}
}

@ARTICLE{Norberg09,
       author = {{Norberg}, P. and {Baugh}, C.~M. and {Gazta{\~n}aga}, E. and {Croton}, D.~J.},
        title = "{Statistical analysis of galaxy surveys - I. Robust error estimation for two-point clustering statistics}",
      journal = {\mnras},
         year = 2009,
        month = jun,
       volume = {396},
       number = {1},
        pages = {19-38},
          doi = {10.1111/j.1365-2966.2009.14389.x},
archivePrefix = {arXiv},
       eprint = {0810.1885},
 primaryClass = {astro-ph},
       adsurl = {https://ui.adsabs.harvard.edu/abs/2009MNRAS.396...19N}
}

@ARTICLE{Asgari21,
       author = {{Asgari}, Marika and {Lin}, Chieh-An and {Joachimi}, Benjamin and {Giblin}, Benjamin and {Heymans}, Catherine and {Hildebrandt}, Hendrik and {Kannawadi}, Arun and {St{\"o}lzner}, Benjamin and {Tr{\"o}ster}, Tilman and {van den Busch}, Jan Luca and {Wright}, Angus H. and {Bilicki}, Maciej and {Blake}, Chris and {de Jong}, Jelte and {Dvornik}, Andrej and {Erben}, Thomas and {Getman}, Fedor and {Hoekstra}, Henk and {K{\"o}hlinger}, Fabian and {Kuijken}, Konrad and {Miller}, Lance and {Radovich}, Mario and {Schneider}, Peter and {Shan}, HuanYuan and {Valentijn}, Edwin},
        title = "{KiDS-1000 cosmology: Cosmic shear constraints and comparison between two point statistics}",
      journal = {\aap},
         year = 2021,
        month = jan,
       volume = {645},
          eid = {A104},
        pages = {A104},
          doi = {10.1051/0004-6361/202039070},
archivePrefix = {arXiv},
       eprint = {2007.15633},
 primaryClass = {astro-ph.CO},
       adsurl = {https://ui.adsabs.harvard.edu/abs/2021A&A...645A.104A}
}

@article{Secco22,
  title = {Dark Energy Survey Year 3 results: Cosmology from cosmic shear and robustness to modeling uncertainty},
  author = {Secco, L. F. and Samuroff, S. and Krause, E. and Jain, B. and Blazek, J. and Raveri, M. and Campos, A. and Amon, A. and Chen, A. and Doux, C. and Choi, A. and Gruen, D. and Bernstein, G. M. and Chang, C. and DeRose, J. and Myles, J. and Fert\'e, A. and Lemos, P. and Huterer, D. and Prat, J. and Troxel, M. A. and MacCrann, N. and Liddle, A. R. and Kacprzak, T. and Fang, X. and S\'anchez, C. and Pandey, S. and Dodelson, S. and Chintalapati, P. and Hoffmann, K. and Alarcon, A. and Alves, O. and Andrade-Oliveira, F. and Baxter, E. J. and Bechtol, K. and Becker, M. R. and Brandao-Souza, A. and Camacho, H. and Carnero Rosell, A. and Carrasco Kind, M. and Cawthon, R. and Cordero, J. P. and Crocce, M. and Davis, C. and Di Valentino, E. and Drlica-Wagner, A. and Eckert, K. and Eifler, T. F. and Elidaiana, M. and Elsner, F. and Elvin-Poole, J. and Everett, S. and Fosalba, P. and Friedrich, O. and Gatti, M. and Giannini, G. and Gruendl, R. A. and Harrison, I. and Hartley, W. G. and Herner, K. and Huang, H. and Huff, E. M. and Jarvis, M. and Jeffrey, N. and Kuropatkin, N. and Leget, P.-F. and Muir, J. and Mccullough, J. and Navarro Alsina, A. and Omori, Y. and Park, Y. and Porredon, A. and Rollins, R. and Roodman, A. and Rosenfeld, R. and Ross, A. J. and Rykoff, E. S. and Sanchez, J. and Sevilla-Noarbe, I. and Sheldon, E. S. and Shin, T. and Troja, A. and Tutusaus, I. and Varga, T. N. and Weaverdyck, N. and Wechsler, R. H. and Yanny, B. and Yin, B. and Zhang, Y. and Zuntz, J. and Abbott, T. M. C. and Aguena, M. and Allam, S. and Annis, J. and Bacon, D. and Bertin, E. and Bhargava, S. and Bridle, S. L. and Brooks, D. and Buckley-Geer, E. and Burke, D. L. and Carretero, J. and Costanzi, M. and da Costa, L. N. and De Vicente, J. and Diehl, H. T. and Dietrich, J. P. and Doel, P. and Ferrero, I. and Flaugher, B. and Frieman, J. and Garc\'{\i}a-Bellido, J. and Gaztanaga, E. and Gerdes, D. W. and Giannantonio, T. and Gschwend, J. and Gutierrez, G. and Hinton, S. R. and Hollowood, D. L. and Honscheid, K. and Hoyle, B. and James, D. J. and Jeltema, T. and Kuehn, K. and Lahav, O. and Lima, M. and Lin, H. and Maia, M. A. G. and Marshall, J. L. and Martini, P. and Melchior, P. and Menanteau, F. and Miquel, R. and Mohr, J. J. and Morgan, R. and Ogando, R. L. C. and Palmese, A. and Paz-Chinch\'on, F. and Petravick, D. and Pieres, A. and Plazas Malag\'on, A. A. and Rodriguez-Monroy, M. and Romer, A. K. and Sanchez, E. and Scarpine, V. and Schubnell, M. and Scolnic, D. and Serrano, S. and Smith, M. and Soares-Santos, M. and Suchyta, E. and Swanson, M. E. C. and Tarle, G. and Thomas, D. and To, C.},
  collaboration = {DES Collaboration},
  journal = {Phys. Rev. D},
  volume = {105},
  issue = {2},
  pages = {023515},
  numpages = {41},
  year = {2022},
  month = {Jan},
  publisher = {American Physical Society},
  doi = {10.1103/PhysRevD.105.023515},
  url = {https://link.aps.org/doi/10.1103/PhysRevD.105.023515}
}

@article{Amon22,
  title = {Dark Energy Survey Year 3 results: Cosmology from cosmic shear and robustness to data calibration},
  author = {Amon, A. and Gruen, D. and Troxel, M. A. and MacCrann, N. and Dodelson, S. and Choi, A. and Doux, C. and Secco, L. F. and Samuroff, S. and Krause, E. and Cordero, J. and Myles, J. and DeRose, J. and Wechsler, R. H. and Gatti, M. and Navarro-Alsina, A. and Bernstein, G. M. and Jain, B. and Blazek, J. and Alarcon, A. and Fert\'e, A. and Lemos, P. and Raveri, M. and Campos, A. and Prat, J. and S\'anchez, C. and Jarvis, M. and Alves, O. and Andrade-Oliveira, F. and Baxter, E. and Bechtol, K. and Becker, M. R. and Bridle, S. L. and Camacho, H. and Carnero Rosell, A. and Carrasco Kind, M. and Cawthon, R. and Chang, C. and Chen, R. and Chintalapati, P. and Crocce, M. and Davis, C. and Diehl, H. T. and Drlica-Wagner, A. and Eckert, K. and Eifler, T. F. and Elvin-Poole, J. and Everett, S. and Fang, X. and Fosalba, P. and Friedrich, O. and Gaztanaga, E. and Giannini, G. and Gruendl, R. A. and Harrison, I. and Hartley, W. G. and Herner, K. and Huang, H. and Huff, E. M. and Huterer, D. and Kuropatkin, N. and Leget, P. and Liddle, A. R. and McCullough, J. and Muir, J. and Pandey, S. and Park, Y. and Porredon, A. and Refregier, A. and Rollins, R. P. and Roodman, A. and Rosenfeld, R. and Ross, A. J. and Rykoff, E. S. and Sanchez, J. and Sevilla-Noarbe, I. and Sheldon, E. and Shin, T. and Troja, A. and Tutusaus, I. and Tutusaus, I. and Varga, T. N. and Weaverdyck, N. and Yanny, B. and Yin, B. and Zhang, Y. and Zuntz, J. and Aguena, M. and Allam, S. and Annis, J. and Bacon, D. and Bertin, E. and Bhargava, S. and Brooks, D. and Buckley-Geer, E. and Burke, D. L. and Carretero, J. and Costanzi, M. and da Costa, L. N. and Pereira, M. E. S. and De Vicente, J. and Desai, S. and Dietrich, J. P. and Doel, P. and Ferrero, I. and Flaugher, B. and Frieman, J. and Garc\'{\i}a-Bellido, J. and Gaztanaga, E. and Gerdes, D. W. and Giannantonio, T. and Gschwend, J. and Gutierrez, G. and Hinton, S. R. and Hollowood, D. L. and Honscheid, K. and Hoyle, B. and James, D. J. and Kron, R. and Kuehn, K. and Lahav, O. and Lima, M. and Lin, H. and Maia, M. A. G. and Marshall, J. L. and Martini, P. and Melchior, P. and Menanteau, F. and Miquel, R. and Mohr, J. J. and Morgan, R. and Ogando, R. L. C. and Palmese, A. and Paz-Chinch\'on, F. and Petravick, D. and Pieres, A. and Romer, A. K. and Sanchez, E. and Scarpine, V. and Schubnell, M. and Serrano, S. and Smith, M. and Soares-Santos, M. and Tarle, G. and Thomas, D. and To, C. and Weller, J.},
  collaboration = {DES Collaboration},
  journal = {Phys. Rev. D},
  volume = {105},
  issue = {2},
  pages = {023514},
  numpages = {43},
  year = {2022},
  month = {Jan},
  publisher = {American Physical Society},
  doi = {10.1103/PhysRevD.105.023514},
  url = {https://link.aps.org/doi/10.1103/PhysRevD.105.023514}
}

@ARTICLE{Eisenstein98,
       author = {{Eisenstein}, Daniel J. and {Hu}, Wayne},
        title = "{Baryonic Features in the Matter Transfer Function}",
      journal = {\apj},
         year = 1998,
        month = mar,
       volume = {496},
       number = {2},
        pages = {605-614},
          doi = {10.1086/305424},
archivePrefix = {arXiv},
       eprint = {astro-ph/9709112},
 primaryClass = {astro-ph},
       adsurl = {https://ui.adsabs.harvard.edu/abs/1998ApJ...496..605E}
}

@ARTICLE{Baldauf13,
       author = {{Baldauf}, Tobias and {Seljak}, Uro{\v{s}} and {Smith}, Robert E. and {Hamaus}, Nico and {Desjacques}, Vincent},
        title = "{Halo stochasticity from exclusion and nonlinear clustering}",
      journal = {\prd},
         year = 2013,
        month = oct,
       volume = {88},
       number = {8},
          eid = {083507},
        pages = {083507},
          doi = {10.1103/PhysRevD.88.083507},
archivePrefix = {arXiv},
       eprint = {1305.2917},
 primaryClass = {astro-ph.CO},
       adsurl = {https://ui.adsabs.harvard.edu/abs/2013PhRvD..88h3507B}
}

@ARTICLE{Bardeen86,
       author = {{Bardeen}, J.~M. and {Bond}, J.~R. and {Kaiser}, N. and {Szalay}, A.~S.},
        title = "{The Statistics of Peaks of Gaussian Random Fields}",
      journal = {\apj},
         year = 1986,
        month = may,
       volume = {304},
        pages = {15},
          doi = {10.1086/164143},
       adsurl = {https://ui.adsabs.harvard.edu/abs/1986ApJ...304...15B}
}

@ARTICLE{Tormen98,
       author = {{Tormen}, Giuseppe},
        title = "{The assembly of matter in galaxy clusters}",
      journal = {\mnras},
         year = 1998,
        month = jun,
       volume = {297},
       number = {2},
        pages = {648-656},
          doi = {10.1046/j.1365-8711.1998.01545.x},
archivePrefix = {arXiv},
       eprint = {astro-ph/9802290},
 primaryClass = {astro-ph},
       adsurl = {https://ui.adsabs.harvard.edu/abs/1998MNRAS.297..648T}
}

@ARTICLE{Giocoli10,
       author = {{Giocoli}, Carlo and {Bartelmann}, Matthias and {Sheth}, Ravi K. and {Cacciato}, Marcello},
        title = "{Halo model description of the non-linear dark matter power spectrum at k >> 1Mpc$^{-1}$}",
      journal = {\mnras},
         year = 2010,
        month = oct,
       volume = {408},
       number = {1},
        pages = {300-313},
          doi = {10.1111/j.1365-2966.2010.17108.x},
archivePrefix = {arXiv},
       eprint = {1003.4740},
 primaryClass = {astro-ph.CO},
       adsurl = {https://ui.adsabs.harvard.edu/abs/2010MNRAS.408..300G}
}

@ARTICLE{Romanello24,
	author = {Romanello, M. and {Marulli, F.} and {Moscardini, L.} and {Lesci, G. F.} and {Sartoris, B.} and {Contarini, S.} and {Giocoli, C.} and {Bardelli, S.} and {Busillo, V.} and {Castignani, G.} and {Covone, G.} and {Ingoglia, L.} and {Maturi, M.} and {Puddu, E.} and {Radovich, M.} and {Roncarelli, M.} and {Sereno, M.}},
	title = {AMICO galaxy clusters in KiDS-DR3: Cosmological constraints from the angular power spectrum and correlation function},
	DOI= "10.1051/0004-6361/202348305",
	url= "https://doi.org/10.1051/0004-6361/202348305",
	journal = {A\&A},
	year = 2024,
	volume = 682,
	pages = "A72",
}

@ARTICLE{Abbott22,
       author = {{Abbott}, T.~M.~C. and {Aguena}, M. and {Alarcon}, A. and {Allam}, S. and {Alves}, O. and {Amon}, A. and {Andrade-Oliveira}, F. and {Annis}, J. and {Avila}, S. and {Bacon}, D. and {Baxter}, E. and {Bechtol}, K. and {Becker}, M.~R. and {Bernstein}, G.~M. and {Bhargava}, S. and {Birrer}, S. and {Blazek}, J. and {Brandao-Souza}, A. and {Bridle}, S.~L. and {Brooks}, D. and {Buckley-Geer}, E. and {Burke}, D.~L. and {Camacho}, H. and {Campos}, A. and {Carnero Rosell}, A. and {Carrasco Kind}, M. and {Carretero}, J. and {Castander}, F.~J. and {Cawthon}, R. and {Chang}, C. and {Chen}, A. and {Chen}, R. and {Choi}, A. and {Conselice}, C. and {Cordero}, J. and {Costanzi}, M. and {Crocce}, M. and {da Costa}, L.~N. and {da Silva Pereira}, M.~E. and {Davis}, C. and {Davis}, T.~M. and {De Vicente}, J. and {DeRose}, J. and {Desai}, S. and {Di Valentino}, E. and {Diehl}, H.~T. and {Dietrich}, J.~P. and {Dodelson}, S. and {Doel}, P. and {Doux}, C. and {Drlica-Wagner}, A. and {Eckert}, K. and {Eifler}, T.~F. and {Elsner}, F. and {Elvin-Poole}, J. and {Everett}, S. and {Evrard}, A.~E. and {Fang}, X. and {Farahi}, A. and {Fernandez}, E. and {Ferrero}, I. and {Fert{\'e}}, A. and {Fosalba}, P. and {Friedrich}, O. and {Frieman}, J. and {Garc{\'\i}a-Bellido}, J. and {Gatti}, M. and {Gaztanaga}, E. and {Gerdes}, D.~W. and {Giannantonio}, T. and {Giannini}, G. and {Gruen}, D. and {Gruendl}, R.~A. and {Gschwend}, J. and {Gutierrez}, G. and {Harrison}, I. and {Hartley}, W.~G. and {Herner}, K. and {Hinton}, S.~R. and {Hollowood}, D.~L. and {Honscheid}, K. and {Hoyle}, B. and {Huff}, E.~M. and {Huterer}, D. and {Jain}, B. and {James}, D.~J. and {Jarvis}, M. and {Jeffrey}, N. and {Jeltema}, T. and {Kovacs}, A. and {Krause}, E. and {Kron}, R. and {Kuehn}, K. and {Kuropatkin}, N. and {Lahav}, O. and {Leget}, P. -F. and {Lemos}, P. and {Liddle}, A.~R. and {Lidman}, C. and {Lima}, M. and {Lin}, H. and {MacCrann}, N. and {Maia}, M.~A.~G. and {Marshall}, J.~L. and {Martini}, P. and {McCullough}, J. and {Melchior}, P. and {Mena-Fern{\'a}ndez}, J. and {Menanteau}, F. and {Miquel}, R. and {Mohr}, J.~J. and {Morgan}, R. and {Muir}, J. and {Myles}, J. and {Nadathur}, S. and {Navarro-Alsina}, A. and {Nichol}, R.~C. and {Ogando}, R.~L.~C. and {Omori}, Y. and {Palmese}, A. and {Pandey}, S. and {Park}, Y. and {Paz-Chinch{\'o}n}, F. and {Petravick}, D. and {Pieres}, A. and {Plazas Malag{\'o}n}, A.~A. and {Porredon}, A. and {Prat}, J. and {Raveri}, M. and {Rodriguez-Monroy}, M. and {Rollins}, R.~P. and {Romer}, A.~K. and {Roodman}, A. and {Rosenfeld}, R. and {Ross}, A.~J. and {Rykoff}, E.~S. and {Samuroff}, S. and {S{\'a}nchez}, C. and {Sanchez}, E. and {Sanchez}, J. and {Sanchez Cid}, D. and {Scarpine}, V. and {Schubnell}, M. and {Scolnic}, D. and {Secco}, L.~F. and {Serrano}, S. and {Sevilla-Noarbe}, I. and {Sheldon}, E. and {Shin}, T. and {Smith}, M. and {Soares-Santos}, M. and {Suchyta}, E. and {Swanson}, M.~E.~C. and {Tabbutt}, M. and {Tarle}, G. and {Thomas}, D. and {To}, C. and {Troja}, A. and {Troxel}, M.~A. and {Tucker}, D.~L. and {Tutusaus}, I. and {Varga}, T.~N. and {Walker}, A.~R. and {Weaverdyck}, N. and {Wechsler}, R. and {Weller}, J. and {Yanny}, B. and {Yin}, B. and {Zhang}, Y. and {Zuntz}, J. and {DES Collaboration}},
        title = "{Dark Energy Survey Year 3 results: Cosmological constraints from galaxy clustering and weak lensing}",
      journal = {\prd},
         year = 2022,
        month = jan,
       volume = {105},
       number = {2},
          eid = {023520},
        pages = {023520},
          doi = {10.1103/PhysRevD.105.023520},
archivePrefix = {arXiv},
       eprint = {2105.13549},
 primaryClass = {astro-ph.CO},
       adsurl = {https://ui.adsabs.harvard.edu/abs/2022PhRvD.105b3520A}
}

@ARTICLE{Heymans21,
       author = {{Heymans}, Catherine and {Tr{\"o}ster}, Tilman and {Asgari}, Marika and {Blake}, Chris and {Hildebrandt}, Hendrik and {Joachimi}, Benjamin and {Kuijken}, Konrad and {Lin}, Chieh-An and {S{\'a}nchez}, Ariel G. and {van den Busch}, Jan Luca and {Wright}, Angus H. and {Amon}, Alexandra and {Bilicki}, Maciej and {de Jong}, Jelte and {Crocce}, Martin and {Dvornik}, Andrej and {Erben}, Thomas and {Fortuna}, Maria Cristina and {Getman}, Fedor and {Giblin}, Benjamin and {Glazebrook}, Karl and {Hoekstra}, Henk and {Joudaki}, Shahab and {Kannawadi}, Arun and {K{\"o}hlinger}, Fabian and {Lidman}, Chris and {Miller}, Lance and {Napolitano}, Nicola R. and {Parkinson}, David and {Schneider}, Peter and {Shan}, HuanYuan and {Valentijn}, Edwin A. and {Verdoes Kleijn}, Gijs and {Wolf}, Christian},
        title = "{KiDS-1000 Cosmology: Multi-probe weak gravitational lensing and spectroscopic galaxy clustering constraints}",
      journal = {\aap},
         year = 2021,
        month = feb,
       volume = {646},
          eid = {A140},
        pages = {A140},
          doi = {10.1051/0004-6361/202039063},
archivePrefix = {arXiv},
       eprint = {2007.15632},
 primaryClass = {astro-ph.CO},
       adsurl = {https://ui.adsabs.harvard.edu/abs/2021A&A...646A.140H}
}

@ARTICLE{Springel01,
       author = {{Springel}, Volker and {Yoshida}, Naoki and {White}, Simon D.~M.},
        title = "{GADGET: a code for collisionless and gasdynamical cosmological simulations}",
      journal = {\na},
         year = 2001,
        month = apr,
       volume = {6},
       number = {2},
        pages = {79-117},
          doi = {10.1016/S1384-1076(01)00042-2},
archivePrefix = {arXiv},
       eprint = {astro-ph/0003162},
 primaryClass = {astro-ph},
       adsurl = {https://ui.adsabs.harvard.edu/abs/2001NewA....6...79S}
}

@ARTICLE{Springel05,
       author = {{Springel}, Volker},
        title = "{The cosmological simulation code GADGET-2}",
      journal = {\mnras},
         year = 2005,
        month = dec,
       volume = {364},
       number = {4},
        pages = {1105-1134},
          doi = {10.1111/j.1365-2966.2005.09655.x},
archivePrefix = {arXiv},
       eprint = {astro-ph/0505010},
 primaryClass = {astro-ph},
       adsurl = {https://ui.adsabs.harvard.edu/abs/2005MNRAS.364.1105S}
}

@ARTICLE{Vogelsberger20,
       author = {{Vogelsberger}, Mark and {Marinacci}, Federico and {Torrey}, Paul and {Puchwein}, Ewald},
        title = "{Cosmological simulations of galaxy formation}",
      journal = {Nature Reviews Physics},
         year = 2020,
        month = jan,
       volume = {2},
       number = {1},
        pages = {42-66},
          doi = {10.1038/s42254-019-0127-2},
archivePrefix = {arXiv},
       eprint = {1909.07976},
 primaryClass = {astro-ph.GA},
       adsurl = {https://ui.adsabs.harvard.edu/abs/2020NatRP...2...42V}
}

@ARTICLE{Springel18,
       author = {{Springel}, Volker and {Pakmor}, R{\"u}diger and {Pillepich}, Annalisa and {Weinberger}, Rainer and {Nelson}, Dylan and {Hernquist}, Lars and {Vogelsberger}, Mark and {Genel}, Shy and {Torrey}, Paul and {Marinacci}, Federico and {Naiman}, Jill},
        title = "{First results from the IllustrisTNG simulations: matter and galaxy clustering}",
      journal = {\mnras},
         year = 2018,
        month = mar,
       volume = {475},
       number = {1},
        pages = {676-698},
          doi = {10.1093/mnras/stx3304},
archivePrefix = {arXiv},
       eprint = {1707.03397},
 primaryClass = {astro-ph.GA},
       adsurl = {https://ui.adsabs.harvard.edu/abs/2018MNRAS.475..676S}
}

@ARTICLE{To2021,
       author = {{To}, C. and {Krause}, E. and {Rozo}, E. and {Wu}, H. and {Gruen}, D. and {Wechsler}, R.~H. and {Eifler}, T.~F. and {Rykoff}, E.~S. and {Costanzi}, M. and {Becker}, M.~R. and {Bernstein}, G.~M. and {Blazek}, J. and {Bocquet}, S. and {Bridle}, S.~L. and {Cawthon}, R. and {Choi}, A. and {Crocce}, M. and {Davis}, C. and {DeRose}, J. and {Drlica-Wagner}, A. and {Elvin-Poole}, J. and {Fang}, X. and {Farahi}, A. and {Friedrich}, O. and {Gatti}, M. and {Gaztanaga}, E. and {Giannantonio}, T. and {Hartley}, W.~G. and {Hoyle}, B. and {Jarvis}, M. and {MacCrann}, N. and {McClintock}, T. and {Miranda}, V. and {Pereira}, M.~E.~S. and {Park}, Y. and {Porredon}, A. and {Prat}, J. and {Rau}, M.~M. and {Ross}, A.~J. and {Samuroff}, S. and {S{\'a}nchez}, C. and {Sevilla-Noarbe}, I. and {Sheldon}, E. and {Troxel}, M.~A. and {Varga}, T.~N. and {Vielzeuf}, P. and {Zhang}, Y. and {Zuntz}, J. and {Abbott}, T.~M.~C. and {Aguena}, M. and {Amon}, A. and {Annis}, J. and {Avila}, S. and {Bertin}, E. and {Bhargava}, S. and {Brooks}, D. and {Burke}, D.~L. and {Carnero Rosell}, A. and {Carrasco Kind}, M. and {Carretero}, J. and {Chang}, C. and {Conselice}, C. and {da Costa}, L.~N. and {Davis}, T.~M. and {Desai}, S. and {Diehl}, H.~T. and {Dietrich}, J.~P. and {Everett}, S. and {Evrard}, A.~E. and {Ferrero}, I. and {Flaugher}, B. and {Fosalba}, P. and {Frieman}, J. and {Garc{\'\i}a-Bellido}, J. and {Gruendl}, R.~A. and {Gutierrez}, G. and {Hinton}, S.~R. and {Hollowood}, D.~L. and {Honscheid}, K. and {Huterer}, D. and {James}, D.~J. and {Jeltema}, T. and {Kron}, R. and {Kuehn}, K. and {Kuropatkin}, N. and {Lima}, M. and {Maia}, M.~A.~G. and {Marshall}, J.~L. and {Menanteau}, F. and {Miquel}, R. and {Morgan}, R. and {Muir}, J. and {Myles}, J. and {Palmese}, A. and {Paz-Chinch{\'o}n}, F. and {Plazas}, A.~A. and {Romer}, A.~K. and {Roodman}, A. and {Sanchez}, E. and {Santiago}, B. and {Scarpine}, V. and {Serrano}, S. and {Smith}, M. and {Suchyta}, E. and {Swanson}, M.~E.~C. and {Tarle}, G. and {Thomas}, D. and {Tucker}, D.~L. and {Weller}, J. and {Wester}, W. and {Wilkinson}, R.~D. and {DES Collaboration}},
        title = "{Dark Energy Survey Year 1 Results: Cosmological Constraints from Cluster Abundances, Weak Lensing, and Galaxy Correlations}",
      journal = {\prl},
         year = 2021,
        month = apr,
       volume = {126},
       number = {14},
          eid = {141301},
        pages = {141301},
          doi = {10.1103/PhysRevLett.126.141301},
archivePrefix = {arXiv},
       eprint = {2010.01138},
 primaryClass = {astro-ph.CO},
       adsurl = {https://ui.adsabs.harvard.edu/abs/2021PhRvL.126n1301T}
}

@ARTICLE{Bullock17,
       author = {{Bullock}, James S. and {Boylan-Kolchin}, Michael},
        title = "{Small-Scale Challenges to the {\ensuremath{\Lambda}}CDM Paradigm}",
      journal = {\araa},
         year = 2017,
        month = aug,
       volume = {55},
       number = {1},
        pages = {343-387},
          doi = {10.1146/annurev-astro-091916-055313},
archivePrefix = {arXiv},
       eprint = {1707.04256},
 primaryClass = {astro-ph.CO},
       adsurl = {https://ui.adsabs.harvard.edu/abs/2017ARA&A..55..343B}
}

@ARTICLE{Klypin99,
       author = {{Klypin}, Anatoly and {Kravtsov}, Andrey V. and {Valenzuela}, Octavio and {Prada}, Francisco},
        title = "{Where Are the Missing Galactic Satellites?}",
      journal = {\apj},
         year = 1999,
        month = sep,
       volume = {522},
       number = {1},
        pages = {82-92},
          doi = {10.1086/307643},
archivePrefix = {arXiv},
       eprint = {astro-ph/9901240},
 primaryClass = {astro-ph},
       adsurl = {https://ui.adsabs.harvard.edu/abs/1999ApJ...522...82K}
}

@ARTICLE{DelPopolo17,
       author = {{Del Popolo}, Antonino and {Le Delliou}, Morgan},
        title = "{Small Scale Problems of the {\ensuremath{\Lambda}}CDM Model: A Short Review}",
      journal = {Galaxies},
         year = 2017,
        month = feb,
       volume = {5},
       number = {1},
          eid = {17},
        pages = {17},
          doi = {10.3390/galaxies5010017},
archivePrefix = {arXiv},
       eprint = {1606.07790},
 primaryClass = {astro-ph.CO},
       adsurl = {https://ui.adsabs.harvard.edu/abs/2017Galax...5...17D}
}

@ARTICLE{Kilbinger15,
       author = {{Kilbinger}, Martin},
        title = "{Cosmology with cosmic shear observations: a review}",
      journal = {Reports on Progress in Physics},
         year = 2015,
        month = jul,
       volume = {78},
       number = {8},
          eid = {086901},
        pages = {086901},
          doi = {10.1088/0034-4885/78/8/086901},
archivePrefix = {arXiv},
       eprint = {1411.0115},
 primaryClass = {astro-ph.CO},
       adsurl = {https://ui.adsabs.harvard.edu/abs/2015RPPh...78h6901K}
}

@ARTICLE{Hildebrandt20,
       author = {{Hildebrandt}, H. and {K{\"o}hlinger}, F. and {van den Busch}, J.~L. and {Joachimi}, B. and {Heymans}, C. and {Kannawadi}, A. and {Wright}, A.~H. and {Asgari}, M. and {Blake}, C. and {Hoekstra}, H. and {Joudaki}, S. and {Kuijken}, K. and {Miller}, L. and {Morrison}, C.~B. and {Tr{\"o}ster}, T. and {Amon}, A. and {Archidiacono}, M. and {Brieden}, S. and {Choi}, A. and {de Jong}, J.~T.~A. and {Erben}, T. and {Giblin}, B. and {Mead}, A. and {Peacock}, J.~A. and {Radovich}, M. and {Schneider}, P. and {Sif{\'o}n}, C. and {Tewes}, M.},
        title = "{KiDS+VIKING-450: Cosmic shear tomography with optical and infrared data}",
      journal = {\aap},
         year = 2020,
        month = jan,
       volume = {633},
          eid = {A69},
        pages = {A69},
          doi = {10.1051/0004-6361/201834878},
archivePrefix = {arXiv},
       eprint = {1812.06076},
 primaryClass = {astro-ph.CO},
       adsurl = {https://ui.adsabs.harvard.edu/abs/2020A&A...633A..69H}
}

@ARTICLE{Weinberg15,
       author = {{Weinberg}, David H. and {Bullock}, James S. and {Governato}, Fabio and {Kuzio de Naray}, Rachel and {Peter}, Annika H.~G.},
        title = "{Cold dark matter: Controversies on small scales}",
      journal = {Proceedings of the National Academy of Science},
         year = 2015,
        month = oct,
       volume = {112},
       number = {40},
        pages = {12249-12255},
          doi = {10.1073/pnas.1308716112},
archivePrefix = {arXiv},
       eprint = {1306.0913},
 primaryClass = {astro-ph.CO},
       adsurl = {https://ui.adsabs.harvard.edu/abs/2015PNAS..11212249W}
}

@ARTICLE{Strigari13,
       author = {{Strigari}, Louis E.},
        title = "{Galactic searches for dark matter}",
      journal = {\physrep},
         year = 2013,
        month = oct,
       volume = {531},
       number = {1},
        pages = {1-88},
          doi = {10.1016/j.physrep.2013.05.004},
archivePrefix = {arXiv},
       eprint = {1211.7090},
 primaryClass = {astro-ph.CO},
       adsurl = {https://ui.adsabs.harvard.edu/abs/2013PhR...531....1S}
}

@ARTICLE{Boylan-Kolchin11,
       author = {{Boylan-Kolchin}, Michael and {Bullock}, James S. and {Kaplinghat}, Manoj},
        title = "{Too big to fail? The puzzling darkness of massive Milky Way subhaloes}",
      journal = {\mnras},
         year = 2011,
        month = jul,
       volume = {415},
       number = {1},
        pages = {L40-L44},
          doi = {10.1111/j.1745-3933.2011.01074.x},
archivePrefix = {arXiv},
       eprint = {1103.0007},
 primaryClass = {astro-ph.CO},
       adsurl = {https://ui.adsabs.harvard.edu/abs/2011MNRAS.415L..40B}
}

@ARTICLE{Navarro96,
       author = {{Navarro}, Julio F. and {Frenk}, Carlos S. and {White}, Simon D.~M.},
        title = "{The Structure of Cold Dark Matter Halos}",
      journal = {\apj},
         year = 1996,
        month = may,
       volume = {462},
        pages = {563},
          doi = {10.1086/177173},
archivePrefix = {arXiv},
       eprint = {astro-ph/9508025},
 primaryClass = {astro-ph},
       adsurl = {https://ui.adsabs.harvard.edu/abs/1996ApJ...462..563N}
}

@ARTICLE{Yue18,
       author = {{Yue}, B. and {Castellano}, M. and {Ferrara}, A. and {Fontana}, A. and {Merlin}, E. and {Amor{\'\i}n}, R. and {Grazian}, A. and {M{\'a}rmol-Queralto}, E. and {Micha{\l}owski}, M.~J. and {Mortlock}, A. and {Paris}, D. and {Parsa}, S. and {Pilo}, S. and {Santini}, P. and {Di Criscienzo}, M.},
        title = "{On the Faint End of the Galaxy Luminosity Function in the Epoch of Reionization: Updated Constraints from the HST Frontier Fields}",
      journal = {\apj},
         year = 2018,
        month = dec,
       volume = {868},
       number = {2},
          eid = {115},
        pages = {115},
          doi = {10.3847/1538-4357/aae77f},
archivePrefix = {arXiv},
       eprint = {1711.05130},
 primaryClass = {astro-ph.GA},
       adsurl = {https://ui.adsabs.harvard.edu/abs/2018ApJ...868..115Y}
}

@ARTICLE{Castellano16,
       author = {{Castellano}, M. and {Yue}, B. and {Ferrara}, A. and {Merlin}, E. and {Fontana}, A. and {Amor{\'\i}n}, R. and {Grazian}, A. and {M{\'a}rmol-Queralto}, E. and {Micha{\l}owski}, M.~J. and {Mortlock}, A. and {Paris}, D. and {Parsa}, S. and {Pilo}, S. and {Santini}, P.},
        title = "{Constraints on Photoionization Feedback from Number Counts of Ultra-faint High-redshift Galaxies in the Frontier Fields}",
      journal = {\apjl},
         year = 2016,
        month = jun,
       volume = {823},
       number = {2},
          eid = {L40},
        pages = {L40},
          doi = {10.3847/2041-8205/823/2/L40},
archivePrefix = {arXiv},
       eprint = {1605.01524},
 primaryClass = {astro-ph.GA},
       adsurl = {https://ui.adsabs.harvard.edu/abs/2016ApJ...823L..40C}
}

@ARTICLE{Brooks13,
       author = {{Brooks}, Alyson M. and {Kuhlen}, Michael and {Zolotov}, Adi and {Hooper}, Dan},
        title = "{A Baryonic Solution to the Missing Satellites Problem}",
      journal = {\apj},
         year = 2013,
        month = mar,
       volume = {765},
       number = {1},
          eid = {22},
        pages = {22},
          doi = {10.1088/0004-637X/765/1/22},
archivePrefix = {arXiv},
       eprint = {1209.5394},
 primaryClass = {astro-ph.CO},
       adsurl = {https://ui.adsabs.harvard.edu/abs/2013ApJ...765...22B}
}

@ARTICLE{Busha10,
       author = {{Busha}, Michael T. and {Alvarez}, Marcelo A. and {Wechsler}, Risa H. and {Abel}, Tom and {Strigari}, Louis E.},
        title = "{The Impact of Inhomogeneous Reionization on the Satellite Galaxy Population of the Milky Way}",
      journal = {\apj},
         year = 2010,
        month = feb,
       volume = {710},
       number = {1},
        pages = {408-420},
          doi = {10.1088/0004-637X/710/1/408},
archivePrefix = {arXiv},
       eprint = {0901.3553},
 primaryClass = {astro-ph.CO},
       adsurl = {https://ui.adsabs.harvard.edu/abs/2010ApJ...710..408B}
}

@ARTICLE{Dayal15,
       author = {{Dayal}, Pratika and {Mesinger}, Andrei and {Pacucci}, Fabio},
        title = "{Early Galaxy Formation in Warm Dark Matter Cosmologies}",
      journal = {\apj},
         year = 2015,
        month = jun,
       volume = {806},
       number = {1},
          eid = {67},
        pages = {67},
          doi = {10.1088/0004-637X/806/1/67},
archivePrefix = {arXiv},
       eprint = {1408.1102},
 primaryClass = {astro-ph.GA},
       adsurl = {https://ui.adsabs.harvard.edu/abs/2015ApJ...806...67D}
}

@ARTICLE{Menci18,
       author = {{Menci}, N. and {Grazian}, A. and {Lamastra}, A. and {Calura}, F. and {Castellano}, M. and {Santini}, P.},
        title = "{Galaxy Formation in Sterile Neutrino Dark Matter Models}",
      journal = {\apj},
         year = 2018,
        month = feb,
       volume = {854},
       number = {1},
          eid = {1},
        pages = {1},
          doi = {10.3847/1538-4357/aaa773},
archivePrefix = {arXiv},
       eprint = {1801.03697},
 primaryClass = {astro-ph.CO},
       adsurl = {https://ui.adsabs.harvard.edu/abs/2018ApJ...854....1M}
}

@ARTICLE{Menci16,
       author = {{Menci}, N. and {Grazian}, A. and {Castellano}, M. and {Sanchez}, N.~G.},
        title = "{A Stringent Limit on the Warm Dark Matter Particle Masses from the Abundance of z = 6 Galaxies in the Hubble Frontier Fields}",
      journal = {\apjl},
         year = 2016,
        month = jul,
       volume = {825},
       number = {1},
          eid = {L1},
        pages = {L1},
          doi = {10.3847/2041-8205/825/1/L1},
archivePrefix = {arXiv},
       eprint = {1606.02530},
 primaryClass = {astro-ph.CO},
       adsurl = {https://ui.adsabs.harvard.edu/abs/2016ApJ...825L...1M}
}

@ARTICLE{Romanello21,
       author = {{Romanello}, Massimiliano and {Menci}, Nicola and {Castellano}, Marco},
        title = "{The Epoch of Reionization in Warm Dark Matter Scenarios}",
      journal = {Universe},
         year = 2021,
        month = sep,
       volume = {7},
       number = {10},
          eid = {365},
        pages = {365},
          doi = {10.3390/universe7100365},
archivePrefix = {arXiv},
       eprint = {2110.05262},
 primaryClass = {astro-ph.CO},
       adsurl = {https://ui.adsabs.harvard.edu/abs/2021Univ....7..365R}
}

@ARTICLE{Corasaniti17,
       author = {{Corasaniti}, P.~S. and {Agarwal}, S. and {Marsh}, D.~J.~E. and {Das}, S.},
        title = "{Constraints on dark matter scenarios from measurements of the galaxy luminosity function at high redshifts}",
      journal = {\prd},
         year = 2017,
        month = apr,
       volume = {95},
       number = {8},
          eid = {083512},
        pages = {083512},
          doi = {10.1103/PhysRevD.95.083512},
archivePrefix = {arXiv},
       eprint = {1611.05892},
 primaryClass = {astro-ph.CO},
       adsurl = {https://ui.adsabs.harvard.edu/abs/2017PhRvD..95h3512C}
}

@ARTICLE{Carucci19,
       author = {{Carucci}, Isabella P. and {Corasaniti}, Pier-Stefano},
        title = "{Cosmic reionization history and dark matter scenarios}",
      journal = {\prd},
         year = 2019,
        month = jan,
       volume = {99},
       number = {2},
          eid = {023518},
        pages = {023518},
          doi = {10.1103/PhysRevD.99.023518},
archivePrefix = {arXiv},
       eprint = {1811.07904},
 primaryClass = {astro-ph.CO},
       adsurl = {https://ui.adsabs.harvard.edu/abs/2019PhRvD..99b3518C}
}

@ARTICLE{Dayal17,
       author = {{Dayal}, Pratika and {Choudhury}, Tirthankar Roy and {Bromm}, Volker and {Pacucci}, Fabio},
        title = "{Reionization and Galaxy Formation in Warm Dark Matter Cosmologies}",
      journal = {\apj},
         year = 2017,
        month = feb,
       volume = {836},
       number = {1},
          eid = {16},
        pages = {16},
          doi = {10.3847/1538-4357/836/1/16},
archivePrefix = {arXiv},
       eprint = {1501.02823},
 primaryClass = {astro-ph.CO},
       adsurl = {https://ui.adsabs.harvard.edu/abs/2017ApJ...836...16D}
}

@ARTICLE{Tulin18,
       author = {{Tulin}, Sean and {Yu}, Hai-Bo},
        title = "{Dark matter self-interactions and small scale structure}",
      journal = {\physrep},
         year = 2018,
        month = feb,
       volume = {730},
        pages = {1-57},
          doi = {10.1016/j.physrep.2017.11.004},
archivePrefix = {arXiv},
       eprint = {1705.02358},
 primaryClass = {hep-ph},
       adsurl = {https://ui.adsabs.harvard.edu/abs/2018PhR...730....1T}
}

@ARTICLE{Adhikari22,
       author = {{Adhikari}, Susmita and {Banerjee}, Arka and {Boddy}, Kimberly K. and {Cyr-Racine}, Francis-Yan and {Desmond}, Harry and {Dvorkin}, Cora and {Jain}, Bhuvnesh and {Kahlhoefer}, Felix and {Kaplinghat}, Manoj and {Nierenberg}, Anna and {Peter}, Annika H.~G. and {Robertson}, Andrew and {Sakstein}, Jeremy and {Zavala}, Jes{\'u}s},
        title = "{Astrophysical tests of dark matter self-interactions}",
      journal = {Reviews of Modern Physics},
         year = 2025,
        month = oct,
       volume = {97},
       number = {4},
          eid = {045004},
        pages = {045004},
          doi = {10.1103/m2vm-59y3},
archivePrefix = {arXiv},
       eprint = {2207.10638},
 primaryClass = {astro-ph.CO},
       adsurl = {https://ui.adsabs.harvard.edu/abs/2025RvMP...97d5004A}
}

@ARTICLE{White78,
       author = {{White}, S.~D.~M. and {Rees}, M.~J.},
        title = "{Core condensation in heavy halos: a two-stage theory for galaxy formation and clustering.}",
      journal = {\mnras},
         year = 1978,
        month = may,
       volume = {183},
        pages = {341-358},
          doi = {10.1093/mnras/183.3.341},
       adsurl = {https://ui.adsabs.harvard.edu/abs/1978MNRAS.183..341W}
}

@ARTICLE{Weinberger18,
       author = {{Weinberger}, Rainer and {Springel}, Volker and {Pakmor}, R{\"u}diger and {Nelson}, Dylan and {Genel}, Shy and {Pillepich}, Annalisa and {Vogelsberger}, Mark and {Marinacci}, Federico and {Naiman}, Jill and {Torrey}, Paul and {Hernquist}, Lars},
        title = "{Supermassive black holes and their feedback effects in the IllustrisTNG simulation}",
      journal = {\mnras},
         year = 2018,
        month = sep,
       volume = {479},
       number = {3},
        pages = {4056-4072},
          doi = {10.1093/mnras/sty1733},
archivePrefix = {arXiv},
       eprint = {1710.04659},
 primaryClass = {astro-ph.GA},
       adsurl = {https://ui.adsabs.harvard.edu/abs/2018MNRAS.479.4056W}
}

@ARTICLE{Smith18,
       author = {{Smith}, Matthew C. and {Sijacki}, Debora and {Shen}, Sijing},
        title = "{Supernova feedback in numerical simulations of galaxy formation: separating physics from numerics}",
      journal = {\mnras},
         year = 2018,
        month = jul,
       volume = {478},
       number = {1},
        pages = {302-331},
          doi = {10.1093/mnras/sty994},
archivePrefix = {arXiv},
       eprint = {1709.03515},
 primaryClass = {astro-ph.GA},
       adsurl = {https://ui.adsabs.harvard.edu/abs/2018MNRAS.478..302S}
}

@ARTICLE{Hopkins12,
       author = {{Hopkins}, Philip F. and {Quataert}, Eliot and {Murray}, Norman},
        title = "{Stellar feedback in galaxies and the origin of galaxy-scale winds}",
      journal = {\mnras},
         year = 2012,
        month = apr,
       volume = {421},
       number = {4},
        pages = {3522-3537},
          doi = {10.1111/j.1365-2966.2012.20593.x},
archivePrefix = {arXiv},
       eprint = {1110.4638},
 primaryClass = {astro-ph.CO},
       adsurl = {https://ui.adsabs.harvard.edu/abs/2012MNRAS.421.3522H}
}

@ARTICLE{Scharre24,
       author = {{Scharr{\'e}}, Lucie and {Sorini}, Daniele and {Dav{\'e}}, Romeel},
        title = "{The effects of stellar and AGN feedback on the cosmic star formation history in the SIMBA simulations}",
      journal = {\mnras},
         year = 2024,
        month = oct,
       volume = {534},
       number = {1},
        pages = {361-383},
          doi = {10.1093/mnras/stae2098},
archivePrefix = {arXiv},
       eprint = {2404.07252},
 primaryClass = {astro-ph.GA},
       adsurl = {https://ui.adsabs.harvard.edu/abs/2024MNRAS.534..361S}
}

@ARTICLE{Chisari18,
       author = {{Chisari}, N.~E. and {Richardson}, M.~L.~A. and {Devriendt}, J. and {Dubois}, Y. and {Schneider}, A. and {Le Brun}, A.~M.~C. and {Beckmann}, R.~S. and {Peirani}, S. and {Slyz}, A. and {Pichon}, C.},
        title = "{The impact of baryons on the matter power spectrum from the Horizon-AGN cosmological hydrodynamical simulation}",
      journal = {\mnras},
         year = 2018,
        month = nov,
       volume = {480},
       number = {3},
        pages = {3962-3977},
          doi = {10.1093/mnras/sty2093},
archivePrefix = {arXiv},
       eprint = {1801.08559},
 primaryClass = {astro-ph.CO},
       adsurl = {https://ui.adsabs.harvard.edu/abs/2018MNRAS.480.3962C}
}

@ARTICLE{Despali25,
       author = {{Despali}, Giulia and {Moscardini}, Lauro and {Nelson}, Dylan and {Pillepich}, Annalisa and {Springel}, Volker and {Vogelsberger}, Mark},
        title = "{Introducing the AIDA-TNG project: Galaxy formation in alternative dark matter models}",
      journal = {\aap},
         year = 2025,
        month = may,
       volume = {697},
          eid = {A213},
        pages = {A213},
          doi = {10.1051/0004-6361/202553836},
archivePrefix = {arXiv},
       eprint = {2501.12439},
 primaryClass = {astro-ph.CO},
       adsurl = {https://ui.adsabs.harvard.edu/abs/2025A&A...697A.213D}
}

@ARTICLE{Pillepich2018,
       author = {{Pillepich}, Annalisa and {Nelson}, Dylan and {Hernquist}, Lars and {Springel}, Volker and {Pakmor}, R{\"u}diger and {Torrey}, Paul and {Weinberger}, Rainer and {Genel}, Shy and {Naiman}, Jill P. and {Marinacci}, Federico and {Vogelsberger}, Mark},
        title = "{First results from the IllustrisTNG simulations: the stellar mass content of groups and clusters of galaxies}",
      journal = {\mnras},
         year = 2018,
        month = mar,
       volume = {475},
       number = {1},
        pages = {648-675},
          doi = {10.1093/mnras/stx3112},
archivePrefix = {arXiv},
       eprint = {1707.03406},
 primaryClass = {astro-ph.GA},
       adsurl = {https://ui.adsabs.harvard.edu/abs/2018MNRAS.475..648P}
}

@ARTICLE{Pillepich18,
       author = {{Pillepich}, Annalisa and {Springel}, Volker and {Nelson}, Dylan and {Genel}, Shy and {Naiman}, Jill and {Pakmor}, R{\"u}diger and {Hernquist}, Lars and {Torrey}, Paul and {Vogelsberger}, Mark and {Weinberger}, Rainer and {Marinacci}, Federico},
        title = "{Simulating galaxy formation with the IllustrisTNG model}",
      journal = {\mnras},
         year = 2018,
        month = jan,
       volume = {473},
       number = {3},
        pages = {4077-4106},
          doi = {10.1093/mnras/stx2656},
archivePrefix = {arXiv},
       eprint = {1703.02970},
 primaryClass = {astro-ph.GA},
       adsurl = {https://ui.adsabs.harvard.edu/abs/2018MNRAS.473.4077P}
}

@ARTICLE{Planck16,
       author = {{Planck Collaboration} and {Ade}, P.~A.~R. and {Aghanim}, N. and {Arnaud}, M. and {Ashdown}, M. and {Aumont}, J. and {Baccigalupi}, C. and {Banday}, A.~J. and {Barreiro}, R.~B. and {Bartlett}, J.~G. and {Bartolo}, N. and {Battaner}, E. and {Battye}, R. and {Benabed}, K. and {Beno{\^\i}t}, A. and {Benoit-L{\'e}vy}, A. and {Bernard}, J. -P. and {Bersanelli}, M. and {Bielewicz}, P. and {Bock}, J.~J. and {Bonaldi}, A. and {Bonavera}, L. and {Bond}, J.~R. and {Borrill}, J. and {Bouchet}, F.~R. and {Bucher}, M. and {Burigana}, C. and {Butler}, R.~C. and {Calabrese}, E. and {Cardoso}, J. -F. and {Catalano}, A. and {Challinor}, A. and {Chamballu}, A. and {Chary}, R. -R. and {Chiang}, H.~C. and {Christensen}, P.~R. and {Church}, S. and {Clements}, D.~L. and {Colombi}, S. and {Colombo}, L.~P.~L. and {Combet}, C. and {Comis}, B. and {Couchot}, F. and {Coulais}, A. and {Crill}, B.~P. and {Curto}, A. and {Cuttaia}, F. and {Danese}, L. and {Davies}, R.~D. and {Davis}, R.~J. and {de Bernardis}, P. and {de Rosa}, A. and {de Zotti}, G. and {Delabrouille}, J. and {D{\'e}sert}, F. -X. and {Diego}, J.~M. and {Dolag}, K. and {Dole}, H. and {Donzelli}, S. and {Dor{\'e}}, O. and {Douspis}, M. and {Ducout}, A. and {Dupac}, X. and {Efstathiou}, G. and {Elsner}, F. and {En{\ss}lin}, T.~A. and {Eriksen}, H.~K. and {Falgarone}, E. and {Fergusson}, J. and {Finelli}, F. and {Forni}, O. and {Frailis}, M. and {Fraisse}, A.~A. and {Franceschi}, E. and {Frejsel}, A. and {Galeotta}, S. and {Galli}, S. and {Ganga}, K. and {Giard}, M. and {Giraud-H{\'e}raud}, Y. and {Gjerl{\o}w}, E. and {Gonz{\'a}lez-Nuevo}, J. and {G{\'o}rski}, K.~M. and {Gratton}, S. and {Gregorio}, A. and {Gruppuso}, A. and {Gudmundsson}, J.~E. and {Hansen}, F.~K. and {Hanson}, D. and {Harrison}, D.~L. and {Henrot-Versill{\'e}}, S. and {Hern{\'a}ndez-Monteagudo}, C. and {Herranz}, D. and {Hildebrandt}, S.~R. and {Hivon}, E. and {Hobson}, M. and {Holmes}, W.~A. and {Hornstrup}, A. and {Hovest}, W. and {Huffenberger}, K.~M. and {Hurier}, G. and {Jaffe}, A.~H. and {Jaffe}, T.~R. and {Jones}, W.~C. and {Juvela}, M. and {Keih{\"a}nen}, E. and {Keskitalo}, R. and {Kisner}, T.~S. and {Kneissl}, R. and {Knoche}, J. and {Kunz}, M. and {Kurki-Suonio}, H. and {Lagache}, G. and {L{\"a}hteenm{\"a}ki}, A. and {Lamarre}, J. -M. and {Lasenby}, A. and {Lattanzi}, M. and {Lawrence}, C.~R. and {Leonardi}, R. and {Lesgourgues}, J. and {Levrier}, F. and {Liguori}, M. and {Lilje}, P.~B. and {Linden-V{\o}rnle}, M. and {L{\'o}pez-Caniego}, M. and {Lubin}, P.~M. and {Mac{\'\i}as-P{\'e}rez}, J.~F. and {Maggio}, G. and {Maino}, D. and {Mandolesi}, N. and {Mangilli}, A. and {Maris}, M. and {Martin}, P.~G. and {Mart{\'\i}nez-Gonz{\'a}lez}, E. and {Masi}, S. and {Matarrese}, S. and {McGehee}, P. and {Meinhold}, P.~R. and {Melchiorri}, A. and {Melin}, J. -B. and {Mendes}, L. and {Mennella}, A. and {Migliaccio}, M. and {Mitra}, S. and {Miville-Desch{\^e}nes}, M. -A. and {Moneti}, A. and {Montier}, L. and {Morgante}, G. and {Mortlock}, D. and {Moss}, A. and {Munshi}, D. and {Murphy}, J.~A. and {Naselsky}, P. and {Nati}, F. and {Natoli}, P. and {Netterfield}, C.~B. and {N{\o}rgaard-Nielsen}, H.~U. and {Noviello}, F. and {Novikov}, D. and {Novikov}, I. and {Oxborrow}, C.~A. and {Paci}, F. and {Pagano}, L. and {Pajot}, F. and {Paoletti}, D. and {Partridge}, B. and {Pasian}, F. and {Patanchon}, G. and {Pearson}, T.~J. and {Perdereau}, O. and {Perotto}, L. and {Perrotta}, F. and {Pettorino}, V. and {Piacentini}, F. and {Piat}, M. and {Pierpaoli}, E. and {Pietrobon}, D. and {Plaszczynski}, S. and {Pointecouteau}, E. and {Polenta}, G. and {Popa}, L. and {Pratt}, G.~W. and {Pr{\'e}zeau}, G. and {Prunet}, S. and {Puget}, J. -L. and {Rachen}, J.~P. and {Rebolo}, R. and {Reinecke}, M. and {Remazeilles}, M. and {Renault}, C. and {Renzi}, A. and {Ristorcelli}, I. and {Rocha}, G. and {Roman}, M. and {Rosset}, C. and {Rossetti}, M. and {Roudier}, G. and {Rubi{\~n}o-Mart{\'\i}n}, J.~A. and {Rusholme}, B. and {Sandri}, M.},
        title = "{Planck 2015 results. XXIV. Cosmology from Sunyaev-Zeldovich cluster counts}",
      journal = {\aap},
         year = 2016,
        month = sep,
       volume = {594},
          eid = {A24},
        pages = {A24},
          doi = {10.1051/0004-6361/201525833},
archivePrefix = {arXiv},
       eprint = {1502.01597},
 primaryClass = {astro-ph.CO},
       adsurl = {https://ui.adsabs.harvard.edu/abs/2016A&A...594A..24P}
}

@ARTICLE{Springel2001,
       author = {{Springel}, Volker and {White}, Simon D.~M. and {Tormen}, Giuseppe and {Kauffmann}, Guinevere},
        title = "{Populating a cluster of galaxies - I. Results at z=0}",
      journal = {\mnras},
         year = 2001,
        month = dec,
       volume = {328},
       number = {3},
        pages = {726-750},
          doi = {10.1046/j.1365-8711.2001.04912.x},
archivePrefix = {arXiv},
       eprint = {astro-ph/0012055},
 primaryClass = {astro-ph},
       adsurl = {https://ui.adsabs.harvard.edu/abs/2001MNRAS.328..726S}
}

@ARTICLE{Viel13,
       author = {{Viel}, Matteo and {Becker}, George D. and {Bolton}, James S. and {Haehnelt}, Martin G.},
        title = "{Warm dark matter as a solution to the small scale crisis: New constraints from high redshift Lyman-{\ensuremath{\alpha}} forest data}",
      journal = {\prd},
         year = 2013,
        month = aug,
       volume = {88},
       number = {4},
          eid = {043502},
        pages = {043502},
          doi = {10.1103/PhysRevD.88.043502},
archivePrefix = {arXiv},
       eprint = {1306.2314},
 primaryClass = {astro-ph.CO},
       adsurl = {https://ui.adsabs.harvard.edu/abs/2013PhRvD..88d3502V}
}

@ARTICLE{Dekker22,
       author = {{Dekker}, Ariane and {Ando}, Shin'ichiro and {Correa}, Camila A. and {Ng}, Kenny C.~Y.},
        title = "{Warm dark matter constraints using Milky Way satellite observations and subhalo evolution modeling}",
      journal = {\prd},
         year = 2022,
        month = dec,
       volume = {106},
       number = {12},
          eid = {123026},
        pages = {123026},
          doi = {10.1103/PhysRevD.106.123026},
archivePrefix = {arXiv},
       eprint = {2111.13137},
 primaryClass = {astro-ph.CO},
       adsurl = {https://ui.adsabs.harvard.edu/abs/2022PhRvD.106l3026D}
}

@ARTICLE{Villasenor23,
       author = {{Villasenor}, Bruno and {Robertson}, Brant and {Madau}, Piero and {Schneider}, Evan},
        title = "{New constraints on warm dark matter from the Lyman-{\ensuremath{\alpha}} forest power spectrum}",
      journal = {\prd},
         year = 2023,
        month = jul,
       volume = {108},
       number = {2},
          eid = {023502},
        pages = {023502},
          doi = {10.1103/PhysRevD.108.023502},
archivePrefix = {arXiv},
       eprint = {2209.14220},
 primaryClass = {astro-ph.CO},
       adsurl = {https://ui.adsabs.harvard.edu/abs/2023PhRvD.108b3502V}
}

@ARTICLE{Correa21,
       author = {{Correa}, Camila A.},
        title = "{Constraining velocity-dependent self-interacting dark matter with the Milky Way's dwarf spheroidal galaxies}",
      journal = {\mnras},
         year = 2021,
        month = may,
       volume = {503},
       number = {1},
        pages = {920-937},
          doi = {10.1093/mnras/stab506},
archivePrefix = {arXiv},
       eprint = {2007.02958},
 primaryClass = {astro-ph.GA},
       adsurl = {https://ui.adsabs.harvard.edu/abs/2021MNRAS.503..920C}
}

@ARTICLE{Zheng05,
       author = {{Zheng}, Zheng and {Berlind}, Andreas A. and {Weinberg}, David H. and {Benson}, Andrew J. and {Baugh}, Carlton M. and {Cole}, Shaun and {Dav{\'e}}, Romeel and {Frenk}, Carlos S. and {Katz}, Neal and {Lacey}, Cedric G.},
        title = "{Theoretical Models of the Halo Occupation Distribution: Separating Central and Satellite Galaxies}",
      journal = {\apj},
         year = 2005,
        month = nov,
       volume = {633},
       number = {2},
        pages = {791-809},
          doi = {10.1086/466510},
archivePrefix = {arXiv},
       eprint = {astro-ph/0408564},
 primaryClass = {astro-ph},
       adsurl = {https://ui.adsabs.harvard.edu/abs/2005ApJ...633..791Z}
}

@ARTICLE{Asgari23,
       author = {{Asgari}, Marika and {Mead}, Alexander J. and {Heymans}, Catherine},
        title = "{The halo model for cosmology: a pedagogical review}",
      journal = {The Open Journal of Astrophysics},
         year = 2023,
        month = nov,
       volume = {6},
          eid = {39},
        pages = {39},
          doi = {10.21105/astro.2303.08752},
archivePrefix = {arXiv},
       eprint = {2303.08752},
 primaryClass = {astro-ph.CO},
       adsurl = {https://ui.adsabs.harvard.edu/abs/2023OJAp....6E..39A}
}

@ARTICLE{Piscionere15,
       author = {{Piscionere}, Jennifer A. and {Berlind}, Andreas A. and {McBride}, Cameron K. and {Scoccimarro}, Rom{\'a}n},
        title = "{The Spatial Distribution of Satellite Galaxies within Halos: Measuring the Very Small Scale Angular Clustering of SDSS Galaxies}",
      journal = {\apj},
         year = 2015,
        month = jun,
       volume = {806},
       number = {1},
          eid = {125},
        pages = {125},
          doi = {10.1088/0004-637X/806/1/125},
archivePrefix = {arXiv},
       eprint = {1407.6740},
 primaryClass = {astro-ph.GA},
       adsurl = {https://ui.adsabs.harvard.edu/abs/2015ApJ...806..125P}
}

@ARTICLE{Skibba15,
       author = {{Skibba}, Ramin A. and {Coil}, Alison L. and {Mendez}, Alexander J. and {Blanton}, Michael R. and {Bray}, Aaron D. and {Cool}, Richard J. and {Eisenstein}, Daniel J. and {Guo}, Hong and {Miyaji}, Takamitsu and {Moustakas}, John and {Zhu}, Guangtun},
        title = "{Dark Matter Halo Models of Stellar Mass-dependent Galaxy Clustering in PRIMUS+DEEP2 at 0.2>z>1.2}",
      journal = {\apj},
         year = 2015,
        month = jul,
       volume = {807},
       number = {2},
          eid = {152},
        pages = {152},
          doi = {10.1088/0004-637X/807/2/152},
archivePrefix = {arXiv},
       eprint = {1503.00731},
 primaryClass = {astro-ph.CO},
       adsurl = {https://ui.adsabs.harvard.edu/abs/2015ApJ...807..152S}
}

@article{Watson10,
doi = {10.1088/0004-637X/709/1/115},
url = {https://dx.doi.org/10.1088/0004-637X/709/1/115},
year = {2009},
month = {dec},
publisher = {The American Astronomical Society},
volume = {709},
number = {1},
pages = {115},
author = {Watson, Douglas F. and Berlind, Andreas A. and McBride, Cameron K. and Masjedi, Morad},
title = {MODELING THE VERY SMALL SCALE CLUSTERING OF LUMINOUS RED GALAXIES},
journal = {The Astrophysical Journal},

}

@ARTICLE{More09,
       author = {{More}, Surhud and {van den Bosch}, Frank C. and {Cacciato}, Marcello and {Mo}, H.~J. and {Yang}, Xiaohu and {Li}, Ran},
        title = "{Satellite kinematics - II. The halo mass-luminosity relation of central galaxies in SDSS}",
      journal = {\mnras},
         year = 2009,
        month = jan,
       volume = {392},
       number = {2},
        pages = {801-816},
          doi = {10.1111/j.1365-2966.2008.14095.x},
archivePrefix = {arXiv},
       eprint = {0807.4532},
 primaryClass = {astro-ph},
       adsurl = {https://ui.adsabs.harvard.edu/abs/2009MNRAS.392..801M}
}

@ARTICLE{Zehavi05,
       author = {{Zehavi}, Idit and {Zheng}, Zheng and {Weinberg}, David H. and {Frieman}, Joshua A. and {Berlind}, Andreas A. and {Blanton}, Michael R. and {Scoccimarro}, Roman and {Sheth}, Ravi K. and {Strauss}, Michael A. and {Kayo}, Issha and {Suto}, Yasushi and {Fukugita}, Masataka and {Nakamura}, Osamu and {Bahcall}, Neta A. and {Brinkmann}, Jon and {Gunn}, James E. and {Hennessy}, Greg S. and {Ivezi{\'c}}, {\v{Z}}eljko and {Knapp}, Gillian R. and {Loveday}, Jon and {Meiksin}, Avery and {Schlegel}, David J. and {Schneider}, Donald P. and {Szapudi}, Istvan and {Tegmark}, Max and {Vogeley}, Michael S. and {York}, Donald G. and {SDSS Collaboration}},
        title = "{The Luminosity and Color Dependence of the Galaxy Correlation Function}",
      journal = {\apj},
         year = 2005,
        month = sep,
       volume = {630},
       number = {1},
        pages = {1-27},
          doi = {10.1086/431891},
archivePrefix = {arXiv},
       eprint = {astro-ph/0408569},
 primaryClass = {astro-ph},
       adsurl = {https://ui.adsabs.harvard.edu/abs/2005ApJ...630....1Z}
}

@ARTICLE{Berlind02,
       author = {{Berlind}, Andreas A. and {Weinberg}, David H.},
        title = "{The Halo Occupation Distribution: Toward an Empirical Determination of the Relation between Galaxies and Mass}",
      journal = {\apj},
         year = 2002,
        month = aug,
       volume = {575},
       number = {2},
        pages = {587-616},
          doi = {10.1086/341469},
archivePrefix = {arXiv},
       eprint = {astro-ph/0109001},
 primaryClass = {astro-ph},
       adsurl = {https://ui.adsabs.harvard.edu/abs/2002ApJ...575..587B}
}

@ARTICLE{Ghigna98,
       author = {{Ghigna}, Sebastiano and {Moore}, Ben and {Governato}, Fabio and {Lake}, George and {Quinn}, Thomas and {Stadel}, Joachim},
        title = "{Dark matter haloes within clusters}",
      journal = {\mnras},
         year = 1998,
        month = oct,
       volume = {300},
       number = {1},
        pages = {146-162},
          doi = {10.1046/j.1365-8711.1998.01918.x},
archivePrefix = {arXiv},
       eprint = {astro-ph/9801192},
 primaryClass = {astro-ph},
       adsurl = {https://ui.adsabs.harvard.edu/abs/1998MNRAS.300..146G}
}

@ARTICLE{Springel08,
       author = {{Springel}, V. and {Wang}, J. and {Vogelsberger}, M. and {Ludlow}, A. and {Jenkins}, A. and {Helmi}, A. and {Navarro}, J.~F. and {Frenk}, C.~S. and {White}, S.~D.~M.},
        title = "{The Aquarius Project: the subhaloes of galactic haloes}",
      journal = {\mnras},
         year = 2008,
        month = dec,
       volume = {391},
       number = {4},
        pages = {1685-1711},
          doi = {10.1111/j.1365-2966.2008.14066.x},
archivePrefix = {arXiv},
       eprint = {0809.0898},
 primaryClass = {astro-ph},
       adsurl = {https://ui.adsabs.harvard.edu/abs/2008MNRAS.391.1685S}
}

@ARTICLE{Despali20,
       author = {{Despali}, Giulia and {Lovell}, Mark and {Vegetti}, Simona and {Crain}, Robert A. and {Oppenheimer}, Benjamin D.},
        title = "{The lensing properties of subhaloes in massive elliptical galaxies in sterile neutrino cosmologies}",
      journal = {\mnras},
         year = 2020,
        month = jan,
       volume = {491},
       number = {1},
        pages = {1295-1310},
          doi = {10.1093/mnras/stz3068},
archivePrefix = {arXiv},
       eprint = {1907.06649},
 primaryClass = {astro-ph.CO},
       adsurl = {https://ui.adsabs.harvard.edu/abs/2020MNRAS.491.1295D}
}

@ARTICLE{Bose17,
       author = {{Bose}, Sownak and {Hellwing}, Wojciech A. and {Frenk}, Carlos S. and {Jenkins}, Adrian and {Lovell}, Mark R. and {Helly}, John C. and {Li}, Baojiu and {Gonzalez-Perez}, Violeta and {Gao}, Liang},
        title = "{Substructure and galaxy formation in the Copernicus Complexio warm dark matter simulations}",
      journal = {\mnras},
         year = 2017,
        month = feb,
       volume = {464},
       number = {4},
        pages = {4520-4533},
          doi = {10.1093/mnras/stw2686},
archivePrefix = {arXiv},
       eprint = {1604.07409},
 primaryClass = {astro-ph.CO},
       adsurl = {https://ui.adsabs.harvard.edu/abs/2017MNRAS.464.4520B}
}

@ARTICLE{Han16,
       author = {{Han}, Jiaxin and {Cole}, Shaun and {Frenk}, Carlos S. and {Jing}, Yipeng},
        title = "{A unified model for the spatial and mass distribution of subhaloes}",
      journal = {\mnras},
         year = 2016,
        month = apr,
       volume = {457},
       number = {2},
        pages = {1208-1223},
          doi = {10.1093/mnras/stv2900},
archivePrefix = {arXiv},
       eprint = {1509.02175},
 primaryClass = {astro-ph.CO},
       adsurl = {https://ui.adsabs.harvard.edu/abs/2016MNRAS.457.1208H}
}

@ARTICLE{Ross13,
       author = {{Ross}, Ashley J. and {Percival}, Will J. and {Carnero}, Aurelio and {Zhao}, Gong-bo and {Manera}, Marc and {Raccanelli}, Alvise and {Aubourg}, Eric and {Bizyaev}, Dmitry and {Brewington}, Howard and {Brinkmann}, J. and {Brownstein}, Joel R. and {Cuesta}, Antonio J. and {da Costa}, Luiz A.~N. and {Eisenstein}, Daniel J. and {Ebelke}, Garrett and {Guo}, Hong and {Hamilton}, Jean-Christophe and {Maga{\~n}a}, Mariana Vargas and {Malanushenko}, Elena and {Malanushenko}, Viktor and {Maraston}, Claudia and {Montesano}, Francesco and {Nichol}, Robert C. and {Oravetz}, Daniel and {Pan}, Kaike and {Prada}, Francisco and {S{\'a}nchez}, Ariel G. and {Samushia}, Lado and {Schlegel}, David J. and {Schneider}, Donald P. and {Seo}, Hee-Jong and {Sheldon}, Alaina and {Simmons}, Audrey and {Snedden}, Stephanie and {Swanson}, Molly E.~C. and {Thomas}, Daniel and {Tinker}, Jeremy L. and {Tojeiro}, Rita and {Zehavi}, Idit},
        title = "{The clustering of galaxies in the SDSS-III DR9 Baryon Oscillation Spectroscopic Survey: constraints on primordial non-Gaussianity}",
      journal = {\mnras},
         year = 2013,
        month = jan,
       volume = {428},
       number = {2},
        pages = {1116-1127},
          doi = {10.1093/mnras/sts094},
archivePrefix = {arXiv},
       eprint = {1208.1491},
 primaryClass = {astro-ph.CO},
       adsurl = {https://ui.adsabs.harvard.edu/abs/2013MNRAS.428.1116R}
}

@ARTICLE{Riquelme23,
       author = {{Riquelme}, Walter and {Avila}, Santiago and {Garc{\'\i}a-Bellido}, Juan and {Porredon}, Anna and {Ferrero}, Ismael and {Chan}, Kwan Chuen and {Rosenfeld}, Rogerio and {Camacho}, Hugo and {Adame}, Adrian G. and {Carnero Rosell}, Aurelio and {Crocce}, Martin and {De Vicente}, Juan and {Eifler}, Tim and {Elvin-Poole}, Jack and {Fang}, Xiao and {Krause}, Elisabeth and {Rodriguez Monroy}, Martin and {Ross}, Ashley J. and {Sanchez}, Eusebio and {Sevilla}, Ignacio},
        title = "{Primordial non-Gaussianity with angular correlation function: integral constraint and validation for DES}",
      journal = {\mnras},
         year = 2023,
        month = jul,
       volume = {523},
       number = {1},
        pages = {603-619},
          doi = {10.1093/mnras/stad1429},
archivePrefix = {arXiv},
       eprint = {2209.07187},
 primaryClass = {astro-ph.CO},
       adsurl = {https://ui.adsabs.harvard.edu/abs/2023MNRAS.523..603R}
}

@ARTICLE{Lovell14,
       author = {{Lovell}, Mark R. and {Frenk}, Carlos S. and {Eke}, Vincent R. and {Jenkins}, Adrian and {Gao}, Liang and {Theuns}, Tom},
        title = "{The properties of warm dark matter haloes}",
      journal = {\mnras},
         year = 2014,
        month = mar,
       volume = {439},
       number = {1},
        pages = {300-317},
          doi = {10.1093/mnras/stt2431},
archivePrefix = {arXiv},
       eprint = {1308.1399},
 primaryClass = {astro-ph.CO},
       adsurl = {https://ui.adsabs.harvard.edu/abs/2014MNRAS.439..300L}
}

@ARTICLE{Kim18,
       author = {{Kim}, Stacy Y. and {Peter}, Annika H.~G. and {Hargis}, Jonathan R.},
        title = "{Missing Satellites Problem: Completeness Corrections to the Number of Satellite Galaxies in the Milky Way are Consistent with Cold Dark Matter Predictions}",
      journal = {\prl},
         year = 2018,
        month = nov,
       volume = {121},
       number = {21},
          eid = {211302},
        pages = {211302},
          doi = {10.1103/PhysRevLett.121.211302},
archivePrefix = {arXiv},
       eprint = {1711.06267},
 primaryClass = {astro-ph.CO},
       adsurl = {https://ui.adsabs.harvard.edu/abs/2018PhRvL.121u1302K}
}

@ARTICLE{Bode2001,
       author = {{Bode}, Paul and {Ostriker}, Jeremiah P. and {Turok}, Neil},
        title = "{Halo Formation in Warm Dark Matter Models}",
      journal = {\apj},
         year = 2001,
        month = jul,
       volume = {556},
       number = {1},
        pages = {93-107},
          doi = {10.1086/321541},
archivePrefix = {arXiv},
       eprint = {astro-ph/0010389},
 primaryClass = {astro-ph},
       adsurl = {https://ui.adsabs.harvard.edu/abs/2001ApJ...556...93B}
}

@ARTICLE{Schneider2012,
       author = {{Schneider}, Aurel and {Smith}, Robert E. and {Macci{\`o}}, Andrea V. and {Moore}, Ben},
        title = "{Non-linear evolution of cosmological structures in warm dark matter models}",
      journal = {\mnras},
         year = 2012,
        month = jul,
       volume = {424},
       number = {1},
        pages = {684-698},
          doi = {10.1111/j.1365-2966.2012.21252.x},
archivePrefix = {arXiv},
       eprint = {1112.0330},
 primaryClass = {astro-ph.CO},
       adsurl = {https://ui.adsabs.harvard.edu/abs/2012MNRAS.424..684S}
}

@article{Viel05,
  title = {Constraining warm dark matter candidates including sterile neutrinos and light gravitinos with WMAP and the Lyman-$\ensuremath{\alpha}$ forest},
  author = {Viel, Matteo and Lesgourgues, Julien and Haehnelt, Martin G. and Matarrese, Sabino and Riotto, Antonio},
  journal = {Phys. Rev. D},
  volume = {71},
  issue = {6},
  pages = {063534},
  numpages = {10},
  year = {2005},
  month = {Mar},
  publisher = {American Physical Society},
  doi = {10.1103/PhysRevD.71.063534},
  url = {https://link.aps.org/doi/10.1103/PhysRevD.71.063534}
}

@ARTICLE{Weinberger17,
       author = {{Weinberger}, Rainer and {Springel}, Volker and {Hernquist}, Lars and {Pillepich}, Annalisa and {Marinacci}, Federico and {Pakmor}, R{\"u}diger and {Nelson}, Dylan and {Genel}, Shy and {Vogelsberger}, Mark and {Naiman}, Jill and {Torrey}, Paul},
        title = "{Simulating galaxy formation with black hole driven thermal and kinetic feedback}",
      journal = {\mnras},
         year = 2017,
        month = mar,
       volume = {465},
       number = {3},
        pages = {3291-3308},
          doi = {10.1093/mnras/stw2944},
archivePrefix = {arXiv},
       eprint = {1607.03486},
 primaryClass = {astro-ph.GA},
       adsurl = {https://ui.adsabs.harvard.edu/abs/2017MNRAS.465.3291W}
}

@ARTICLE{Watson12,
       author = {{Watson}, Douglas F. and {Berlind}, Andreas A. and {McBride}, Cameron K. and {Hogg}, David W. and {Jiang}, Tao},
        title = "{The Extreme Small Scales: Do Satellite Galaxies Trace Dark Matter?}",
      journal = {\apj},
         year = 2012,
        month = apr,
       volume = {749},
       number = {1},
          eid = {83},
        pages = {83},
          doi = {10.1088/0004-637X/749/1/83},
archivePrefix = {arXiv},
       eprint = {1108.1195},
 primaryClass = {astro-ph.CO},
       adsurl = {https://ui.adsabs.harvard.edu/abs/2012ApJ...749...83W}
}

@ARTICLE{Mead15,
       author = {{Mead}, A.~J. and {Peacock}, J.~A. and {Heymans}, C. and {Joudaki}, S. and {Heavens}, A.~F.},
        title = "{An accurate halo model for fitting non-linear cosmological power spectra and baryonic feedback models}",
      journal = {\mnras},
         year = 2015,
        month = dec,
       volume = {454},
       number = {2},
        pages = {1958-1975},
          doi = {10.1093/mnras/stv2036},
archivePrefix = {arXiv},
       eprint = {1505.07833},
 primaryClass = {astro-ph.CO},
       adsurl = {https://ui.adsabs.harvard.edu/abs/2015MNRAS.454.1958M}
}

@ARTICLE{Bhowmick18,
       author = {{Bhowmick}, Aklant K. and {Di Matteo}, Tiziana and {Feng}, Yu and {Lanusse}, Francois},
        title = "{The clustering of z > 7 galaxies: predictions from the BLUETIDES simulation}",
      journal = {\mnras},
         year = 2018,
        month = mar,
       volume = {474},
       number = {4},
        pages = {5393-5405},
          doi = {10.1093/mnras/stx3149},
archivePrefix = {arXiv},
       eprint = {1707.02312},
 primaryClass = {astro-ph.CO},
       adsurl = {https://ui.adsabs.harvard.edu/abs/2018MNRAS.474.5393B}
}

@ARTICLE{Tinker05,
       author = {{Tinker}, Jeremy L. and {Weinberg}, David H. and {Zheng}, Zheng and {Zehavi}, Idit},
        title = "{On the Mass-to-Light Ratio of Large-Scale Structure}",
      journal = {\apj},
         year = 2005,
        month = sep,
       volume = {631},
       number = {1},
        pages = {41-58},
          doi = {10.1086/432084},
archivePrefix = {arXiv},
       eprint = {astro-ph/0411777},
 primaryClass = {astro-ph},
       adsurl = {https://ui.adsabs.harvard.edu/abs/2005ApJ...631...41T}
}

@ARTICLE{vandenBosch13,
       author = {{van den Bosch}, Frank C. and {More}, Surhud and {Cacciato}, Marcello and {Mo}, Houjun and {Yang}, Xiaohu},
        title = "{Cosmological constraints from a combination of galaxy clustering and lensing - I. Theoretical framework}",
      journal = {\mnras},
         year = 2013,
        month = apr,
       volume = {430},
       number = {2},
        pages = {725-746},
          doi = {10.1093/mnras/sts006},
archivePrefix = {arXiv},
       eprint = {1206.6890},
 primaryClass = {astro-ph.CO},
       adsurl = {https://ui.adsabs.harvard.edu/abs/2013MNRAS.430..725V}
}

@ARTICLE{Einasto65,
       author = {{Einasto}, J.},
        title = "{On the Construction of a Composite Model for the Galaxy and on the Determination of the System of Galactic Parameters}",
      journal = {Trudy Astrofizicheskogo Instituta Alma-Ata},
         year = 1965,
        month = jan,
       volume = {5},
        pages = {87-100},
       adsurl = {https://ui.adsabs.harvard.edu/abs/1965TrAlm...5...87E}
}

@ARTICLE{Duffy10,
       author = {{Duffy}, Alan R. and {Schaye}, Joop and {Kay}, Scott T. and {Dalla Vecchia}, Claudio and {Battye}, Richard A. and {Booth}, C.~M.},
        title = "{Impact of baryon physics on dark matter structures: a detailed simulation study of halo density profiles}",
      journal = {\mnras},
         year = 2010,
        month = jul,
       volume = {405},
       number = {4},
        pages = {2161-2178},
          doi = {10.1111/j.1365-2966.2010.16613.x},
archivePrefix = {arXiv},
       eprint = {1001.3447},
 primaryClass = {astro-ph.CO},
       adsurl = {https://ui.adsabs.harvard.edu/abs/2010MNRAS.405.2161D}
}

@ARTICLE{Duffy08,
       author = {{Duffy}, Alan R. and {Schaye}, Joop and {Kay}, Scott T. and {Dalla Vecchia}, Claudio},
        title = "{Dark matter halo concentrations in the Wilkinson Microwave Anisotropy Probe year 5 cosmology}",
      journal = {\mnras},
         year = 2008,
        month = oct,
       volume = {390},
       number = {1},
        pages = {L64-L68},
          doi = {10.1111/j.1745-3933.2008.00537.x},
archivePrefix = {arXiv},
       eprint = {0804.2486},
 primaryClass = {astro-ph},
       adsurl = {https://ui.adsabs.harvard.edu/abs/2008MNRAS.390L..64D}
}
\begin{appendix} 

\end{appendix}

\end{document}